\documentclass{mn2e}
\usepackage{psfig}
\usepackage{mnras_cite}
\input epsfig
\begin{document}

\def \vjec{\vfill\eject}
\def\kmsmpc{\mathrm{km}\ \mathrm{s^{-1}} \mathrm{Mpc^{-1}}}
\def\hmpc{h^{-1} \mathrm{Mpc}}
\def\hkpc{h^{-1} \mathrm{kpc}}
\def\msun{M_\odot}
\def\hmsun{h^{-1} M_\odot}
 \def\chisq{$\chi^2$}
\def \m3{{\rm Mark III}}
\def \etal {{et al.\ }}
\def \cf {{\it cf.\ }}
\def \vs {{\it vs.\ }}
\def \via {{\it via\ }}
\def \ie {{\it i.e.\ } }
\def \eg{{\it e.g.\ }}
\def\kms{\mathrm{km~s^{-1}}}
\def\br{{\bf r}}
\def\bv{{\bf v}}
\def\bV{{\bf V}}
\def \p{\partial}
\def \<{\langle}
\def \>{\rangle}

\def\plotfour#1#2#3#4{\centering \leavevmode
\epsfxsize=.24\columnwidth \epsfbox{#1} \hfil
\epsfxsize=.24\columnwidth \epsfbox{#2} \hfil
\epsfxsize=.24\columnwidth \epsfbox{#3} \hfil
\epsfxsize=.24\columnwidth \epsfbox{#4}}

\def\gsim{~\rlap{$>$}{\lower 1.0ex\hbox{$\sim$}}}
\def\lsim{~\rlap{$<$}{\lower 1.0ex\hbox{$\sim$}}}
\def\d{{\rm d}}
\def\c{{\rm c}}

\newcommand{\pa}{\partial}
\newcommand{\cri}{_{\rm cr}}

\title[The persistence of the universal halo profiles]
{The  persistence of the universal halo profiles}

\author[A. A.~El-Zant]
{Amr A. El-Zant$^{1,2}$  \\
$^1$Mail Code 130-33, California Institute of Technology,
Pasadena, CA 91125, USA\\
$^2$CITA, University of Toronto, Ontario  M5S 3H8, Canada (Present Address)}

\maketitle

\begin{abstract}

Simple simulations suggest that the phase space structure  of haloes
identified in cosmological calculations is invariant  under the dynamics induced by sinking
substructure satellites --- the background expands so as to leave the total distribution unchanged. 
We use a Fokker-Planck formulation to show that there are long lived solutions for  densities 
$\rho \sim r^{\gamma}$ and $-2 \le \gamma \la -1$; indices between $-1$ and $-1.5$ corresponding to the inner
cusps of cosmological haloes, where the coupling is strongest; steeper ones to intermediate radii.
 We recover the exact solutions found by Evans \& Collett (1997); reinterpret them in terms of well 
defined background-satellite interaction; and show that these, and all other solutions, are valid for any  
mass spectrum  of substructure, because the governing equation is linear in their mass weighed phase space 
distribution. If the spatial distribution of substructure  has a milder cusp than the total, the system expands; 
when the background has a milder cusp there is  compression. It is not possible for the individual distributions to retain their original form: 
light particles are driven out of low energy states, being replaced by the sinking massive ones.  If the 
clumps are considered solid, this takes the form of an exponential instability, with characteristic 
timescale of the order of the dynamical friction time,  leading to a low energy cutoff in the 
distribution function of the background and a constant density core.
We show that there are long lived solutions with such a  cutoff. They would correspond to a situation 
whereas the clumps are made of dense baryonic material. When stripping is important, as in the case 
of dissipationless substructure, it is likely that this situation is reversed --- the cutoff is now in the 
clump distribution function. A simple description suggests that this renders equilibria even more long lived. 
In all cases it is possible to find solutions that are long lived from 
the thermodynamical (energy transfer) perspective. In systems without stripping the only truly 
stable solutions however are isothermal spheres, but there are double power law solutions
that may be relevant if stripping is involved. The results in this paper suggest that
halo profiles similar to those found in dissipationless cosmological simulations are approximately invariant
under the interaction induced by the presence of substructure satellites --- a necessary condition
for the observed `universality'.  In addition, the total profile, including baryons, should also be invariant; 
provided the latter are initially in the form of dense clumps, whose distribution follows that of the dark matter.

\end{abstract}



\section{Introduction}
\label{sec:intro}

Collapsed structures identified within the context of  dissipationless 
cosmological simulations are invariably found to exhibit some invariant, or 
universal, form for their density profiles --- irrespective of their mass 
or the epoch at which they are identified (see, e.g., \pcite{navetal04}
and the references therein) --- which, moreover, seems to reflect an invariant 
underlying phase space density distribution (\pcite{tanav01}). These results are 
potentially of fundamental importance; they may signal the existence of a generic 
tendency in initially cold and nearly homogeneous gravitational configurations to 
 evolve towards definite final states, with a consistency  reminiscent  of that characterising 
the approach to a Gibbs distributions in laboratory systems, 
irrespective of the details of the initial conditions or the 
intermediate dynamics.

The existence of invariant final states in systems where the microscopic 
dynamics is known to be time reversible (Newtonian equations in our case)
inevitably echoes the protracted debate pertaining to irreversibility in 
laboratory statistical mechanics, the main conundrum being that such `attractor' like
behaviour is characteristic of dissipative systems. 
 Nevertheless, as is well known, 
dissipative dynamics can be effectively introduced in otherwise reversible systems
by incorporating  a degree of randomness. This can either be achieved by the presence of 
a fluctuating force, representing otherwise intractable interactions, or because the intrinsic 
dynamics is so complicated that tracking its detailed evolution from any given initial 
condition becomes impossible --- the accompanying loss of information has  consequences 
similar to the incorporation of random forces.  
 Systems where such processes are in action may exhibit the required 
macroscopic irreversibility that leads to invariant final states.

Even if, in the cosmological context, the `universality' may be only approximate,
the relatively quaint state of affairs that transpires, rare in astrophysical research,
has generated intense interest.
Discussions dealing with the origin of the identified density distributions
can be broadly divided into two main theses along basic nature (the profiles result from the 
initial collapse: ~\pcite{lohoff01,adi01,hio02})
or nurture (they arise from cumulative effect repeated mergers or 
interactions: \pcite{sywt98,dekev03,dekar03}) lines.
These situations encompass the two contexts alluded to above:
irreversible evolution can either arise from the `violent' internal
evolution involving complicated dynamics of the mean field during the 
initial collapse or merger events; or phenomena arising from 
random encounters between, and dynamical friction forces on, component clumps. 
It is the latter category that concerns us here.

The fact that haloes are {\em always} found to have basically the same forms 
for their profiles suggests that {\em both arguments may be relevant}: 
for, while numerical calculations  simulating the  cold collapse of isolated gravitational systems 
do indeed show that the process  can account for halo profiles (e.g., Hiotelis 2002), a halo's tenure in the context
of a cosmological environment will be punctuated by repeated mergers;  
it will contain substructure, which will be constantly replenished by 
continual infall.  The density distribution needs to be invariant under the 
action of such interactions if the profile is to survive. Evidently, this 
will be the case if the underlying phase space structure is unaffected by the
interactions.

In this paper we endevour to show that this is in fact true; that the presence 
of substructure  leaves the total phase space mass distribution of 
haloes with density distributions similar to those of cosmological-simulation-haloes
(at least approximately) invariant. What needs  to be shown is that  
when dynamical friction acts on clumps (representing  dark matter substructure) 
inside a larger halo, causing them 
to sink towards the centre, the increased density that results from the mass deposition is
precisely balanced by an outflow in the parent halo material; and that mass stripping  from 
clumps does not change that basic conclusion.

Since the evolution necessarily involves macroscopically irreversible behaviour, due to dynamical 
friction and random clump-clump encounters, the formulation presented in this paper may also shed light on the 
process of attainment of the universal profile {\em via} substructure interaction. Though 
this issue is only touched upon briefly here.

As in laboratory statistical mechanics, effective randomness, born of the 
presence of granularity in a system, can be introduced  by considering  a {\em macroscopic} 
transport equation that contains a collision term where this randomness is incorporated.
In the context of gravitational systems the usual approximation involves the Fokker-Planck
equation, extensively used in the study of globular clusters for example (Spitzer 1987).
It has also been invoked in the context of cosmological haloes 
(\pcite{ecoll97,wein01,mabert03}). 
The basic assumption in deriving it from more basic
kinetic equations is that the encounters are weak. This implies that it cannot describe the 
dynamics leading to major 
mergers between a few large components of roughly equal mass.
The assumption however 
is well founded when what is involved is the gravitational interaction of a large number of 
clumps (representing substructure) and their motion though a smooth background medium (representing 
a parent halo). This is the picture we will have in mind in this paper.

In Section~\ref{sec:pasim} we discuss simple numerical experiments illustrating the 
phase space dynamics of such a clump-background system. Following this, in 
Section~\ref{sec:physical}, we tackle the question of correspondence  
between the almost exact phase space  invariance present in these calculations and 
the persistence of density profiles in simulations of the significantly more sophisticated 
cosmological context. We then outline the Fokker-Planck 
formulation used in this paper (Section~\ref{sec:FP}) and apply it to pure power law systems 
(Section~\ref{sec:scalefree}).
Section~\ref{sec:times} discusses the evolution timescales on which the total mass distribution changes,
relative to the dynamical friction timescale it takes for the clumps to sink in.  These are then 
applied to the power law solutions in Section~\ref{sec:timescales}. 
 In Section~\ref{sec:instab}, the evolution of the {\em individual}
components is considered in more detail. The consequences of the solutions obtained 
are outlined  in several sections that follow. 
In particular, Section~\ref{sec:core} discusses the formation of a constant density 
core. This occurs in the background material if no stripping is invoked. The effect of 
stripping is dealt with in Section~\ref{sec:strip}. Both configurations 
correspond to approximate equilibria, with the ones involving stripping  longer lived.

Technical background material is outlined in several 
appendices. Its inclusion was found necessary because of the variety of
conventions, terminology and notation used in the literature of the Fokker-Planck
equation as applied to gravitational systems. Moreover, to the author's knowledge,
derivations of some of the results required for this paper have only been been 
published in the (French language) paper of Henon (1961). Alternative 
derivations are presented.

\section{Numerical experiments}
\label{sec:pasim}

\subsection{Invariance of the phase space distribution}

\label{sec:sim}

\begin{figure*}
\psfig{file = 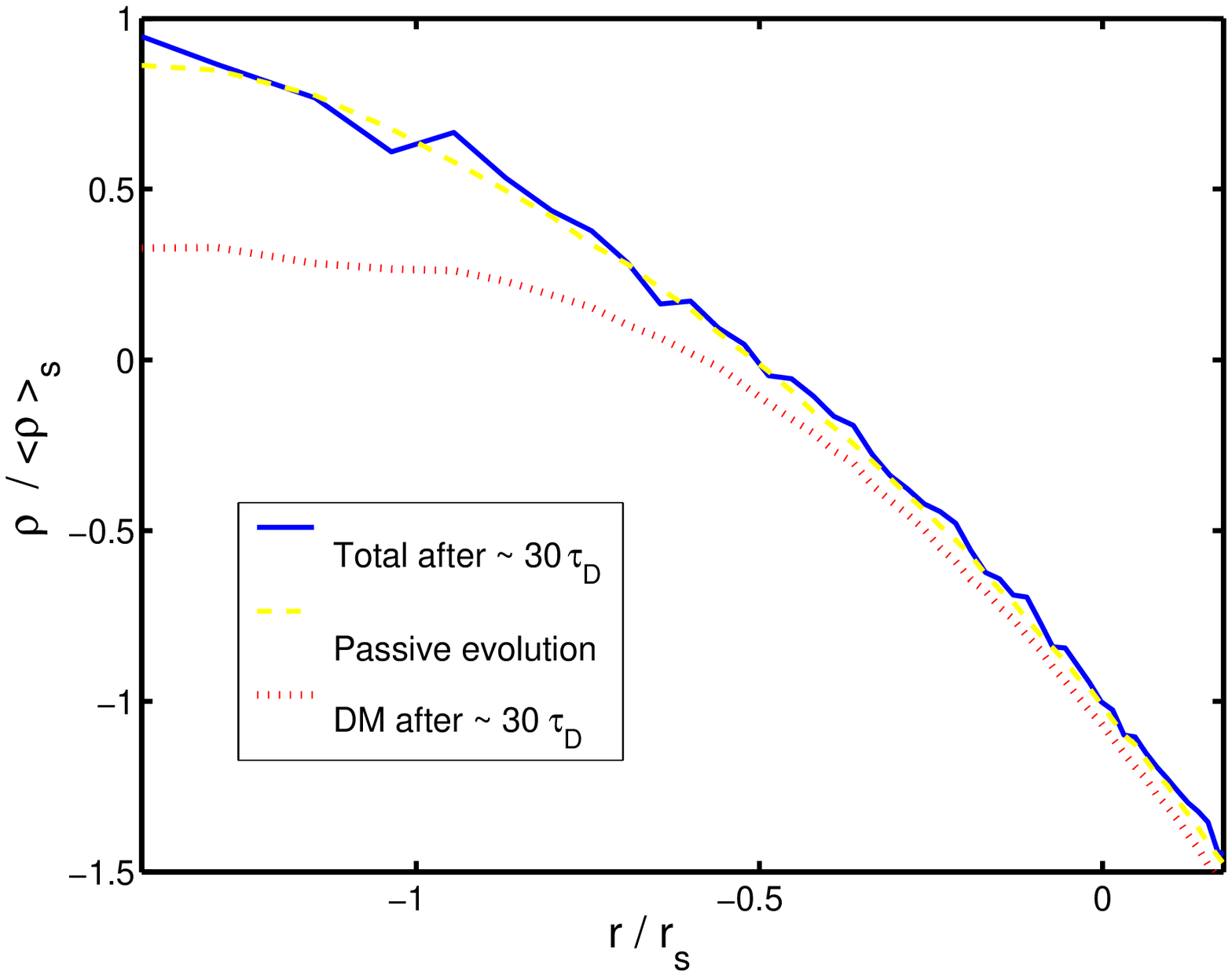,width=88mm}
\psfig{file = 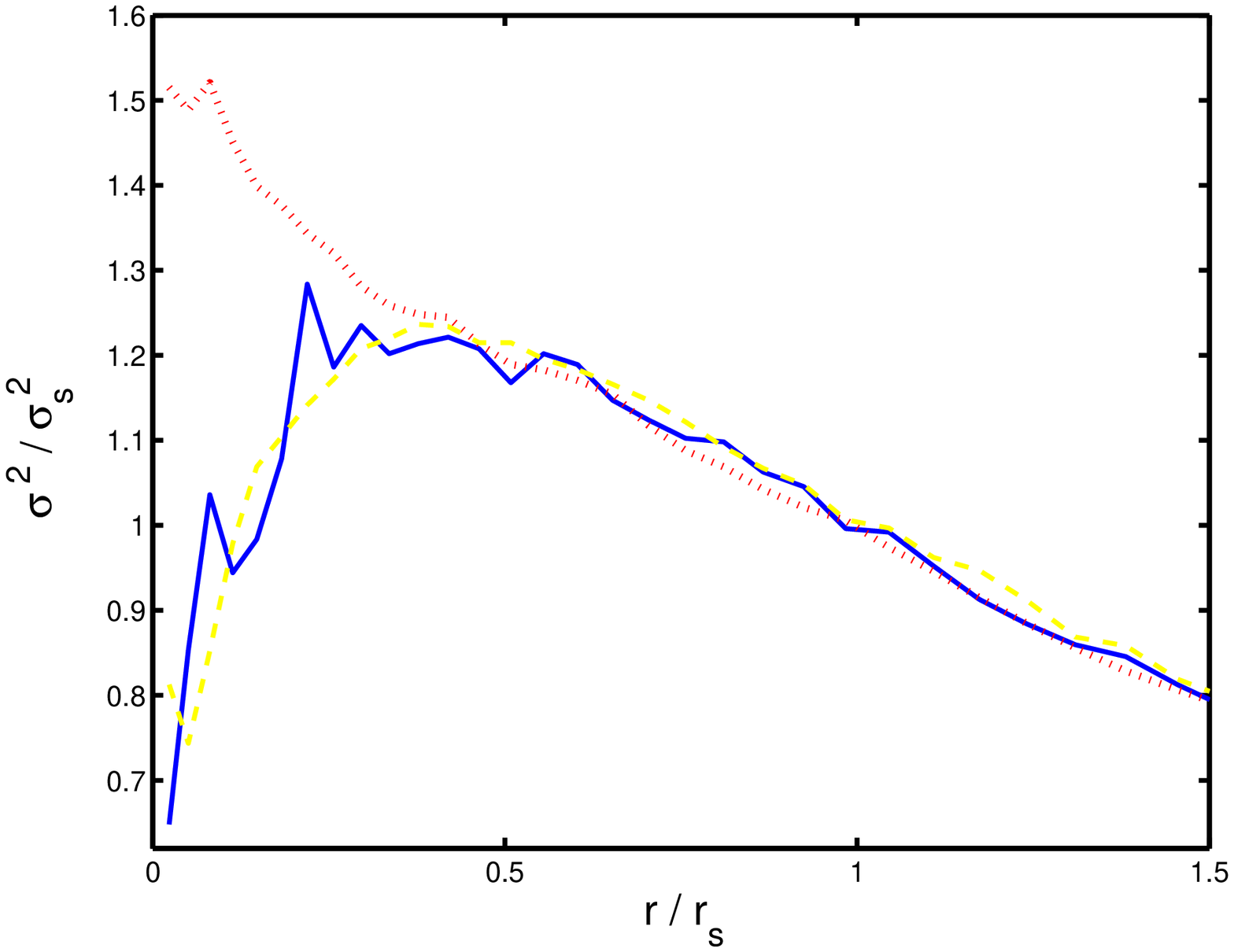,width=87mm}
\caption{Density and velocity distributions of the total and background material composed 
of the lighter particles (denoted by DM). Passive evolution, refers to a system evolved from the same initial conditions,
but with no massive particles. To reduce noise, the  total distribution is 
averaged over three outputs, at $33.5$, $32$ and $30.5$ dynamical times within
the initial scale length $r_s$. The background output, which does not contain high mass
particles and is therefore less noisy, is given at $33.5$ dynamical times,
also at $r_s$. Note  that due to some initial evolution, resulting from the finite size of the 
system, this is slightly different from the scale length  of the distribution shown in the 
figure. Since our comparison involves the passively evolved system, this effect 
is not of major importance.}
\label{fig:sim}
\end{figure*}

Fig.~\ref{fig:sim} shows the density and velocity dispersion distributions of systems of 
massive clumps and smooth background. The simulations are identical to one of  the runs
presented in El-Zant et al. (2004). The clumps are represented by softened point 
particles whose `size' is about $1/40$ of the scale length of the 
Navarro, Frenk \& White (1997) initial density distribution.
 There are 900 000 background (`light') particles through 
which moves a system of  900 clumps with combined mass representing $20 \%$ the 
system's total. They are all started from the same initial 
density and velocity distributions. The system is spherically symmetric with
isotropic velocities, and is evolved using a polar particle mesh code.

As already noted by El-Zant et al (2004), the total density distribution of 
the clump-background system remains virtually unchanged for many dynamical times 
(given in terms of the average density inside the initial 
halo scale length by  $\tau_D = 1/\sqrt{G \< \rho \>_s}$); it is identical, 
except perhaps at the very centre and up to statistical noise, to the passively 
evolved system containing no clumps. 
This statement is also 
true at all times smaller than the timescale shown in the figure (the invariance 
breaks down completely on timescales almost an order of magnitude larger, 
for reasons we discuss in the next subsection).

We have tried different values of the NFW concentration parameter 
$c= r_{vir}/ r_s$. The results presented in this figure refer to a system with 
$c=3.3$. We ran, in addition, simulations  with  of $c= 10$ and $c=33.3$, with almost identical 
results. Different heavy particle masses and fractions were also investigated.
As long as the problem is rescaled in terms of the relevant dynamical and 
dynamical-friction (see next subsection below) timescales, the situation described by Fig.~\ref{fig:sim} 
remains for time intervals of the order of those shown.

 In El-Zant et al (2004) we had interpreted the clump distribution to be composed of baryonic 
material. In that case, the conclusions outlined in that paper still stand:  
the dark matter, made of lighter background particles, is heated and forms a constant density
core;  but the total distribution (interpreted to be composed of baryons and dark matter) 
remains constant. However,  if the clumps are also made of dark matter, 
this invariance implies that the total distribution of dark matter
remains constant.

  A rather remarkable result,
transpiring from Fig.~\ref{fig:sim}, is that it is not only the physical 
density that remains  unchanged, but also the velocity dispersion. 
  The invariance of the velocity distribution is all the more remarkable because 
the initial `temperature inversion' --- the fact that the velocity dispersion decreases 
towards the centre --- is a thermodynamically unstable state of affairs, that should be 
washed out under the action of energy transfer; the system then evolving towards an isothermal 
distribution, as heat flows from hotter to colder 
regions. It so happens, nevertheless, that the `heating' of the lighter particles is precisely 
balanced by a corresponding  `cooling' in the clump distribution, 
so as to keep the total `temperature' constant.

The constancy of the mass and velocity dispersion distributions
in a spherical isotropic 
system, where the distribution function is dependent only on energy,
 must be associated with an invariant  phase space structure. 
In Section~\ref{sec:physical} we argue that {\em this state 
of affairs, so stark in this idealised setting, may remain relevant, resulting in the 
observed persistence of universal profiles, in the highly more complex cosmological context.}
The  remainder of the paper is an attempt to explain and derive the simulations presented here,
and to quantitatively justify this basic contention.

\subsection{Parametrisation in terms of the dynamical friction time}
\label{sec:relax}

\begin{figure}
\psfig{file = 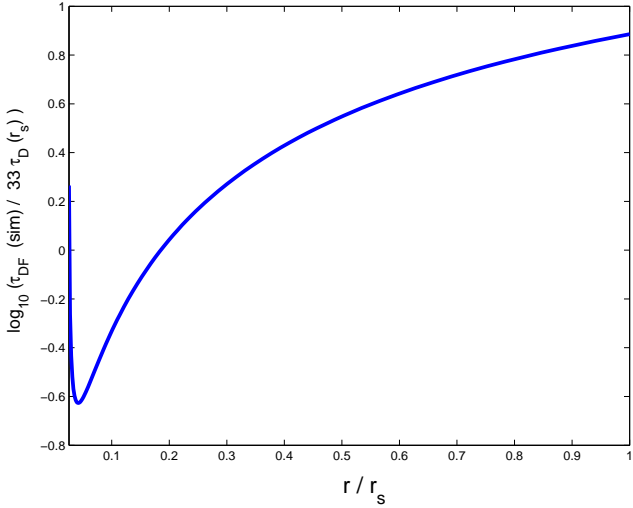,width=85mm}
\caption{Dynamical friction time for an NFW system with $c= 3.3$, relative
to the evolution timescale of the simulation of  Fig.~\ref{fig:sim}, as a function
of radius. It is assumed that the massive particles have `size' $1/40 r_s$}
\label{fig:DFsim}
\end{figure}

  The state of  affairs described by Fig.~\ref{fig:sim} is metastable; in the sense it 
is long-lived, but not  infinitely so. This should be obvious, since once the massive
particles form a self gravitating component, relaxation resulting from their mutual interactions
will result in evolution.

 For timescales of order $200~\tau_D (r_s)$, we have verified 
that the center expands, as energy is transferred from the outer hotter regions. This is in accordance 
to what was found, for example, by Hayashi et al. (2003). Although we have not  performed integrations
for longer intervals, it is expected that it would then proceed towards 
core collapse.

The notion of  solid clumps cannot possibly be useful however in either of  
these regimes; the effect of stripping, which will transform clump material into background, 
ensures that the former can never form an autonomous dynamical system
(cf., Section~\ref{sec:physical}).

We will see later in this paper (Section~\ref{sec:times} and Section~\ref{sec:instab}), 
that the characteristic timescale for the evolution that significantly modifies 
the distribution of the (light and heavy) components, but keeps the total virtually
invariant, is the local dynamical friction time. It will be useful therefore to parametrise 
the  timescale of the simulation just analysed in terms of that quantity.

  To get a rough estimate of the relation between the time the system is evolved in 
Fig.~\ref{fig:sim} and the average  dynamical friction time within a given 
radius, we write the latter as 
(e.g., El-Zant, Shlosman \& Hoffman 2001)~\footnote{Note that this equation 
approximates the density distribution as $\rho \sim 1/r^2$. It also assumes that 
the velocity of the clumps is comparable to the dispersion of the background,
which has a Maxwellian velocity distribution.}
\begin{equation}
\tau_{DF} ({\rm sim})= \frac{E}{\dot{E}} \sim 0.75 \frac{\eta}{\ln \Lambda} \tau_D,
\label{eq:DFsim}
\end{equation}
where $\eta = M ( < r)/ m$ is  of the total system mass inside
radius $r$   divided by the mass of a clump and $\Lambda = \frac{b_{\rm max}}{b_{\rm min}}$ is the 
Coulomb logarithm. We denote this timescale with the label `sim' to 
distinguish it from an analogous  timescale $\tau_{DF}$ that we derive later, 
in a  more rigorous manner, from  diffusion
 coefficients (Section~\ref{sec:times}).

For a system with significant softening, as is the 
case here, $b_{\rm min}$ is determined by the softening length. The maximum
impact parameter can be taken as the radius inside which we are calculating
the average dynamical  friction timescale. We plot in Fig.~\ref{fig:DFsim} 
the dynamical friction time, relative to the 
simulated timescale, for an NFW system with 
$c = r_{\rm vir} / r_s = 3.33$.

Note that the dynamical 
friction timescale is smaller than the simulated one only in the very inner
region, and then only by a factor of a few. (It increases again, because 
the softening dominates. Probably by then however, the formula above becomes entirely
inadequate, the dynamics being  dominated by a non-Newtonian contributions 
to the softened potential). Thus, in terms of dynamical friction time, the system is fairly 
young, despite it being of significant dynamical age (parametrised in terms of $\tau_D$).

To explain the results of the simulation discussed in Section~\ref{sec:sim} therefore, it 
suffices to show that the evolution timescales characterising the change in total phase space 
mass density of a system with central density cusp are much longer than the dynamical friction 
time. Noting also that most of the evolution in the lighter component in Fig.~\ref{fig:sim}
takes place deep inside the $1/r$ cusp, it should be possible, for the purpose of 
explaining the aforementioned results, to approximate the system as a scale free configuration.

\section{Connecting idealised simulations to more complex situations}
\label{sec:physical}

There are two glaring idealisations pertaining  to  the calculations presented in 
the previous section: the heavier bodies are solid softened point particles;
they  also all have the same mass. The advantage, of course, is that the simplified models
serve as to highlight and isolate the interesting phenomenon involving the invariance of the phase space 
distribution --- and identify its origin in the dynamical friction interaction between 
a distinguishable background and a system of massive objects, the number of which 
is large enough so that statistical reasoning may be relevant. But how do these
toy models relate to the actual situation found in realistic simulations of halo  
evolution in the context of CDM cosmologies? In other words, would the discovered 
invariance survive, at least in some approximate sense, the removal of the idealisations?

In the following we attempt to make this connection
 The arguments are only suggestive; though most are broadly  supported by 
the analysis based on the Fokker-Planck formulation presented in subsequent sections of the paper, 
future detailed numerical investigation should also be undertaken.

\subsection{Mass spectrum}

Subhaloes found in cosmological simulations do not all have the same mass. Indeed, 
their differential number  distribution follows the rather steep mass function 
$\frac{d n}{dm} \sim m^{-9/5}$  (e.g., Gao et al 2004c). 
The mass is also a deciding factor as to  how strongly 
a body will interact with a background --- the rate of energy lost by that object
varying as $m^2$.
It would thus appear that in  the assumption of equal masses lies a drastic approximation,
compromising the description of the evolution.

           Nevertheless, it may precisely be the steep
mass spectrum, and the sensitive dependence of the dynamical friction on the mass,
that ensure that the situation described by the single (clump) mass model survives.
For, by virtue of these properties,  a separation of timescales will ensue; the heavier
mass clumps sinking in first, while the lower mass ones are hardly affected by 
dynamical friction. If clump-clump interaction is not of major importance, this initial
evolution will mimic that of the single mass model (clump-clump interaction will 
cause the dynamical evolution of small mass clumps to follow the background material).

          The total distribution of the system just described is also an equilibrium.
Formally,  we show in Section~\ref{sec:FP},  that the fundamental equation deciding the presence of 
an equilibrium is linear with respect to the distribution of massive objects: if a set of 
distributions, possibly corresponding  to different mass species, is a zero of that equation, 
so is their sum (or any linear combination thereof). Thus, one can superpose an equilibrium
distribution made of the evolved system, composed of the heaviest clumps plus background, and 
the distribution of the remaining (non-participating) clumps, which
will remain unchanged if clump-clump interactions can be neglected (or, if such interactions are included,
to various distributions of the remaining clumps).

   There is a  caveat. Detailed invariance will require that the clumps representing
 substructure be initially distributed in a similar manner
to the background. But the viability of a statistical description will
be severely affected by the relatively small number of massive objects in each mass range. 
In particular, the  most massive bodies, which initially contribute most, 
will be represented only by a few specimen. The invariance therefore would only
hold in an ensemble averaged sense. Ma \& Boylan-Kolchin (2004) found that 
substructure may, in individual runs increase or decrease the steepness of 
density profiles; and  cosmological simulations do indeed show 
large scatter in the the profiles of identified haloes (e.g., Klypin et al. 2001).
More work, involving a series of simulations representing a statistical sample of initial 
sustructure distribution and concentrations, in the controlled context of isolated halo calculations, 
is in order.

    If the  clumps come to completely dominate the mass distribution at small radii (as 
happens to the evolved system in Section~\ref{sec:sim}), then the effect of dynamical friction will 
be limited to the most massive specimen --- by the time, at any given energy, the dynamical friction
coupling is strong enough to  affect the next group of clumps, most of the background 
particles that these clumps should couple with would have been removed.
One is then left with a region that can be best described as a system of self gravitating clumps, with 
fluctuating and dissipative forces roughly balancing each other, but where energy transfer (`relaxation')
leads to eventual evolution  (cf. Sections~\ref{sec:theinvariance} and~\ref{sec:core}). 
It is quite likely however that stripping would be effective in transforming much of 
the substructure  into background, before such a situation transpires.  This 
represents another form of cutoff for the dynamical friction-mediated coupling.

   Two question are clearly pertinent. They are concerned with the timescale for stripping 
and its effect on the total phase space mass distribution function.

\subsection{Stripping}
\label{sec:striptease}

 Satellites can no longer be regarded as solid clumps 
if any of the following processes are efficient

\begin{enumerate}

\item
 Particles are removed from the satellites due to pruning by the background
(this includes the possibility of complete dissolution).
\item
 Particles are removed  due to  weak encounters with other satellites.

\item
  Satellites participate in strong mutual encounters or merge.

\end{enumerate}

  The first of these mechanisms does not, to a very good approximation, change the total 
phase space mass distribution function: particles are gently removed from a satellite;
they leave with approximately the same velocity and position as the satellite. Thus, material 
with virtually the same location in phase space is transferred from clump to background, 
the total phase space mass density  remaining nearly constant.

This will also be true of the second mechanism. Weak encounters will
 lead to the gentle removal of particles, they may also facilitate stripping by the main 
halo (cf. Pennerubia \& Benson 2004), both processes through which the total phase space mass 
distribution is closely conserved.

 The situation is different  however if violent encounters and mergers are involved.
It is possible to estimate the timescale associated with such events by averaging 
the clump distribution inside a given radius and invoking an effective collision 
crossection. The envisioned `collisions' will not necessarily lead to mergers (this will depend on their
orbital parameters etc), but we will assume that when two satellites so interact, 
particles in their outer regions can be removed with significantly modified
phase space coordinates..

The collision mean free time for a randomly  moving system of 
$N$ spheres of radius $d$, confined within a spherical volume of 
radius $r$,  can be written as (e.g., Saslaw 1985)
\begin{equation}
\tau_f = \frac{1}{3 N}~\frac{r^2} {d^2}~\tau_{c},
\label{eq:freet}
\end{equation}
where $\tau_c \sim  2 \tau_D$, is the time for an object to make one full crossing of 
the system. 
Clumps of different masses will have different diameters, but one can 
define an effective crossection 
\begin{equation}
\sigma = \frac{\int  \frac{d n}{d m}~d^2~dm}{\int 
\frac{d n}{d m}~d m}. 
\end{equation}
Let the  differential mass function $\frac{d n}{d m} \propto m^{-9/5}$,
and the characteristic radius of a clump be $d = \alpha_1~B~m^{1/3}$, 
where the  parameter $B$ relates  the virial radius and  mass (it depends 
on the redshift and the cosmology; cf. Eq.~A1 of NFW) and $\alpha_1$
is the typical fraction of the virial radius of a clump that survives stripping by 
means of mechanisms i) and ii) in the list above.
In this context we can write
\begin{equation}
\sigma = B^2 \alpha_1^{2}~\frac{\int_{m_{\rm min}}^{m_{\rm max}}  m^{-17/15} dm}
{\int_{m_{\rm min}}^{m_{\rm max}}  m^{-9/5} d m}                                                                         
\approx~\alpha_1^2~B^2~m_{\rm min}^{2/3},
\end{equation}
where we have assumed that the maximal mass in the clump distribution 
is at least a few
orders of magnitude larger than the minimal one. 

Confider a radius $r = \alpha B M_{\rm vir}^{1/3}$,
inside a host halo of  virial mass $M_{\rm vir}$,
and containing $N(r)$ satellites. 
We can  rewrite Eq~(\ref{eq:freet}) in these terms,
replacing  $d^2$ by the effective crossection $\sigma$,
\begin{equation}
\tau_f = \frac{2}{N (r)}~\left(\frac{\alpha}{\alpha_1}\right)^2
~\left(\frac{  M_{\rm vir} } {m_{\rm min}}\right)^{2/3}~~\tau_{\rm cr}.
\label{eq:freetf}
\end{equation}
If  $\alpha_1 = \alpha = 1$, $\tau_f \approx 20~\tau_{\rm cr} \sim 40~\tau_D$ (e.g.,  
if   $N (r_{\rm vir})$ is of order $1000$ and  $m_{\rm min}/M_{\rm vir} \approx 10^{-6}$). 
This represents an average collisional timescale inside the virial radius. It is necessarily
a lower limit, because it effectively assumes that stripping by the host halo, or 
{\em via} weak encounters, is negligible 
(hence $\alpha_1 = 1$).

As one averages over smaller radii (that is $\alpha < 1$), our estimate of 
$\tau_f$ will depend 
on how the ratio $\alpha/\alpha_1$ changes. A simple single power law that 
can be taken as a reasonable representation of the density distribution of 
host and satellites is $\rho \sim 1/r^2$ (we discuss this further 
in Section~\ref{sec:scalefree}). 
In this case, it is easy to show 
(e.g., El-Zant, Kim \& Kamionkowski 2004) that the tidal radius of a satellite 
is proportional to its radial position inside the host halo --- that is 
$\alpha_1 \propto \alpha$. Since $N (r)$ necessarily decreases with $r$, 
this suggests that the above estimate at the virial radius is indeed a
lower limit --- not only there, but at all radii.

During a time interval $\tau_f$, substructure continues to 
be radically modified due to stripping by the host halo; as can be seen from 
Fig.~11 of Taylor \& Babul (2004) for example, where for almost all haloes,
well over $ 90 \%$ of satellite mass is lost on such a timescale 
(note that the radial  period  used there 
$P_{\rm rad} \sim 2 t_{\rm cr} \sim 4 \tau_D$; and 
that the authors do not consider clump-clump  interactions at all. 
Not even weak encounters). It therefore appears unlikely that strong encounters or mergers will
be important at any stage during the evolution ---  haloes being affected 
by gentle stripping on a significantly shorter timescale.

 Finally, note that only a  substructure satellite for which $\tau_{DF} \la 10 \tau_{D}$ 
will be affected by dynamical friction before most of its mass is lost {\em via}
stripping. From Eq.~(\ref{eq:DFsim}) it should have $\eta \ga 0.01$. Thus stripping 
eventually puts an  effective end  to the dynamical friction coupling.

The discussion of the present section suggests that 
a situation whereas sinking clumps would come to dominate the mass distribution
within a certain region (as in Fig.~\ref{fig:sim}) probably would not, in practice, 
arise --- satellites having been significantly stripped long before such a  configuration materialises ---
but  that, nevertheless, the invariance found there will remain if 
\begin{enumerate}

\item

 Stripping is due to gentle removal of particles from clumps 
due to their interaction with background or soft mutual encounters.

\item

Substructure  distributions, properly modified to take into account the 
effect of stripping, also constitute (at least approximate)  equilibria.

\end{enumerate}

 The estimates of the mean free time presented here suggest that
the first condition is satisfied. In Section~\ref{sec:strip} we show that distributions
satisfying condition~2 are indeed excellent approximate equilibria.
In fact, as may already be expected from the discussion above, 
 such configurations are much longer lived, as they 
do not suffer from the problem of relaxation, 
characteristic of a system composed of self gravitating clumps.

\section{Fokker-Planck formulation}
\label{sec:FP}

\subsection{Background}

     The Fokker-Planck equation has been discussed in the context of gravitational
systems in  several textbooks  (Saslaw 1985; Binney \& Tremaine 1987; Sptizer 1987).
The basic assumption is that the perturbations due to the granularity in a system
are weak, local and random. The situation is then analogous to that 
of Brownian motion in a laboratory setting. The dynamical variables therefore undergo 
a random walk,  the amplitude of which is limited by a deterministic force.

Chandrasekhar (1943) reviewed these phenomena, both in the case of laboratory 
systems, where the approximations involved are much more easily justifiable, and 
the gravitational case; for which he derived an explicit formula for the deterministic 
force, that he dubbed dynamical friction. It proved remarkably successful in the 
face of numerical tests, despite the problems associated with justifying the assumptions
of locality and randomness
(e.g. Zaritsky \& White 1988).~\footnote{Higher resolution simulations, testing the associated 
assumptions in detail, are currently overdue nevertheless. For resonant effects, not included in 
the Fokker-Planck analysis, may materialise at higher particle number 
(I thank Martin Weinberg for pointing this out).}

The ratio 
of the amplitude of the fluctuating force causing the random walk to that
of dynamical friction  is determined by the mass of the object in question relative to that
of the perturbers in the background. This implies that more massive particles 
will typically have smaller velocities; in the gravitational context, they will 
sink to the centre of the system. The timescale for  this to happen we call 
the dynamical friction time.  Eq.~(\ref{eq:DFsim}) represents an estimate derived on the basis of the 
Chandrasekhar formula alluded to above. Other evaluations can be directly deduced from the 
Fokker-Planck formulation (Sec.~\ref{sec:times}).

The energy deposited among the population of 
lighter particles causes their velocities to increase and their distribution to 
expand. The principal object of this paper is to show that these processes leave 
the total phase space mass density invariant for timescales significantly longer 
than the dynamical friction time (so as to be negligible on timescales of that order). 
In physical space, that would correspond to the 
expansion of the lighter particle system being almost exactly 
compensated by the inflow of massive clumps.

  Provided that the dynamical friction time  is significantly longer than the orbital 
timescale ($\sim \tau_D$),  the Fokker-Planck equation can be averaged over particle trajectories, 
and  expressed in terms of 
the integrals of motion in the smoothed out potential. The orbit averaged Fokker-Planck 
equation and its properties have been elucidated in detail by Henon (1961), and discussed 
in Binney \& Tremaine (1987) and Spitzer (1987). 
Other works include that of Kohn (1979, 1980) and Merritt (1983), 
the latter being particularly relevant to the situation at hand here.

In the case when the configuration is spherical and has isotopic velocity distribution, as we will assume here,
the six dimensional phase space is also spherically symmetric. The relevant integral of 
motion  is the  energy {\em per unit mass} E, which defines a radial coordinate in that 
space, in terms of which the distribution function and the Fokker-Planck equation may be 
written.~\footnote{Note that $E$  in this paper refers to the energy, and not to the binding energy, which has 
opposite sign, favoured in some of the aforementioned work.} 

We will not be using the Fokker Planck equation directly here, but instead relations 
for the change in mass and energy,  within a given surface $E = {\rm const}$, that can be 
readily derived from it (e.g., Henon 1961 and appendices~\ref{app:mass} and~\ref{app:energy} of the present paper). 
The advantage of this approach is twofolds. First,
the  Fokker-Planck equation itself allows for solutions  with non-vanishing constant
mass flux. These are steady state solutions, even though
they are unphysical, unless there is something to produce or absorb particles  at
the center of  the system, a possibility we do not consider here. The second is 
that steady state solutions of the Fokker Planck equation do not guarantee thermodynamic stability, 
we will need to estimate the effect of energy transfer on the steady state
(cf., Section~\ref{sec:energychange}).

\subsection{Mass change and the fundamental equation for the flux}

\subsubsection{Mass change}

For the mass evolution we have (Appendix~\ref{app:mass})
\begin{equation}
\frac{\p M( < E)} {\p t} =  - F + g \frac{\p q} { \p t},
\label{eq:fok}
\end{equation}
where $g$ is the phase space mass density distribution function --- that is, 
the  amount 
of mass contained in a volume element $\Delta {\bf r} \Delta {\rm v} = (4 \pi)^2 r^2 v^2 dv dr$
is $g \Delta {\bf r} \Delta {\rm v}$.
This equation simply says that the rate of mass change inside the energy surface $E$ 
is determined by the  mass  flux $F$ --- that is, the amount of mass, per unit time, 
 that enters (negative sign)
or exists the phase space volume bounded by the energy surface $E$ ---  
and the mass acquired or lost 
at the edge of this volume because of its variation with time. 
That phase space volume is related  
to the potential; since,  in a spherical phase space,
\begin{equation}
q (E) = (4 \pi)^2 \int r^2  v^2 dv dr =  \frac{16 \pi^2}{3} \int_{0}^{r_{\rm max} (E)} r^2 v^3 dr,
\label{eq:q}
\end{equation}
and $v = \sqrt{2 (E - \phi)}$ (here $r_{\rm max} (E)$ is the maximum radius out to which a particle with 
specific energy $E$ can travel). The `density of states'  $p$,
the area {\em in phase space} of the surface with energy $E$, 
relates to the volume inside that surface by $p =  \frac{\p q}{\p E}$.
Accordingly, the mass enclosed within an energy interval $[E, E+dE]$ 
is $p g dE$; and the total mass within the surface at $E$:
\begin{equation}
M (< E) = \int_{0}^{E} p g dE.
\label{eq:mass}
\end{equation}

 Note that, even when the Fokker Plank flux $F$ is zero,  
the mass distribution may  be time dependent. This state of affairs 
is characteristic, for example, of smooth (collisionless) systems
in configurations out of dynamical equilibrium. In this paper we
assume  that we are dealing not with the merger of a few clumps of
comparable mass, but with a large number of them and a smooth background,
with the combined system having reached a slowly evolving quasi-equilibrium 
state. Any evolution must therefore arise from the flux $F$ describing the collisional
evolution. We will therefore ignore evolution driven solely 
by the second term on the right hand side of~(\ref{eq:fok}).

\subsubsection{General equation for the flux}

  Comparison of the isotropic Fokker Planck equation in flux conservation form 
(Eq.~\ref{eq:FPFC}) and its standard format (e.g., Binney \& Tremaine 1987; Spitzer 1987) yields
\begin{equation}
\frac{\p F}{\p E} = \frac{\p}{\p E} \left(M D_1 - \frac{1}{2} \frac{\p}{\p E} (M D_2) \right);
\end{equation}
from which follows the equation for the flux
\begin{equation}
F = \left(D_E - D_{EE} \frac{\p}{\p E}\right) g,
\label{eq:fluxp}
\end{equation}
where 
\begin{equation}
D_E =  p \< \Delta E \> - \frac{1}{2} \frac{\p}{\p E}~p \< (\Delta E)^2 \>
\label{eq:DE}
\end{equation}
and 
\begin{equation}
D_{EE} = \frac{1}{2}~p\< (\Delta E)^2 \>,
\label{eq:DEE}
\end{equation}
and where $\< \Delta E \> = D_1$ and  $\< (\Delta E)^2 \> = D_2$ are given
 by Eq.~(\ref{eq:D1}) and~(\ref{eq:D2}).

The  Fokker-Planck flux is due to the deterministic and random forces,
acting on the components of the system due to their mutual interactions 
(as discussed in the last subsection).
The changes in energies that result from these forces are represented by the diffusion 
coefficients $D_E$ and $D_{EE}$ respectively. 
It is possible to heuristically 
describe how their forms, and the accompanying expression for the flux, arise.

 When (\ref{eq:DE}) and~(\ref{eq:DEE}) are
inserted into~(\ref{eq:fluxp}), this expression is seen to consist of three 
components. The first term, due to the component $D_E~g$ of~(\ref{eq:fluxp}),  
i.e.~$\< \Delta E \>~p g$, refers to a shell in phase space of 
volume $p~\< \Delta E \>$. On average, due to systematic change in energy $\< \Delta E \>$,
particles  will  traverse such a  shell during a unit time interval.
That is, particles  crossing a  surface at $E = {\rm const}$ in a unit time interval, b 
will come from this shell. The mass inside that shell is the volume multiplied by the 
phase space mass density. Hence the expression $\< \Delta E \>~p g$.

 The random fluctuations characterised by $\< (\Delta E)^2 \>$, as opposed to their weighed gradients  
discussed below, cannot produce a particle flux unless the distribution function has a gradient
--- because particles will be as likely to enter or leave energy surface $E$. 
The RMS energy space `length' traversed in unit time  by a typical particle due to random 
fluctuation is $\delta = \< (\Delta E)^2 \>^{1/2} $. Consider a constant energy surface centered 
on the interval $[E -    \delta , E + \delta]$.
The volume of each of the two shells bounded by $E$ and surfaces at the extremities of 
this interval is  $p  \delta$;
the difference between their average phase space mass density is $\delta \frac{\p g}{\p E}$; and so 
the difference in mass they contain  is 
$p \delta  \times  \delta \frac{\p g}{\p E}$. 
Since the walk in energy space described by  $\< (\Delta E)^2 \>$ is random,
from each of those shells, on average, half the mass would transit in unit                                                                    
time through $E$ (the other half crossing to still higher energies for the
outer  shell, or smaller energies in  the  case of the inner one).   
The magnitude of the resulting mass flux through $E$ is then $\frac{1}{2} p \< (\Delta E)^2 \> \frac{\p g}{\p E}$. 
It is directed outwards (in our convention positive)  when the phase space mass density is a decreasing 
function of energy --- hence  the term $- D_{EE} \frac{\p g}{\p E}$ in~(\ref{eq:fluxp}).

Systematic changes in energy can arise due to the  deterministic dynamical friction
forces; or because the random fluctuations, weighed by the phase space available
at energy $E$, $p = p(E)$, are energy dependent.~\footnote{Note that a flux will arise
even if $\< (\Delta E)^2 \>$ is energy independent; simply because, while particles
will be as likely to cross energy surface $E$ from above or below, if $p$ has a gradient, 
there will be more particles at energy intervals $[E, E+ dE]$ corresponding to larger $p$.}
When this latter component is subtracted from total systematic energy change  $\< \Delta E \> p g$, the result 
represents flux  whose sole source is dynamical friction --- that is, the 
term $D_E~g$ in~(\ref{eq:fluxp}).

\subsubsection{Case of system with two components  carrying  masses $m \gg \mu$}

In such circumstances, the average rates of change of the energy and its square 
are given by~(\ref{eq:deltEmu}), (\ref{eq:deltE}), and~(\ref{eq:deltEE}). 
It is then straightforward to show  that  $D_E = 0$ for the light particles.
For the heavy particles, it consists of the second term of
Eq.~(\ref{eq:deltE}).~\footnote{Note that our definition of $D_E$
contains an extra mass factor as compared with that adopted by Merritt 1983, who
prefers to multiply the term $D_E$ in Eq.~(\ref{eq:fluxp}) by the mass.}
One can also see that $D_{EE}$ is the same for both species, and depends only on the 
distribution of the  heavy masses.

This is to be expected. Dynamical friction on the lighter particles is negligible;
they can also be considered non-interacting, 
their evolution in energy space being determined  by the random 
energy they, on average, gain from the massive particles. For, if these heavier
particles are removed, the remaining system, composed of light particles, 
is collisionless.

The flux in light (background) particles is then
\begin{equation}
F_\mu = -  D_{EE} \frac{\p g_\mu}{\p E}.
\label{eq:flight}
\end{equation}
For the massive particles, on the other hand, we have
\begin{equation}
F_m = \left(D_E  - D_{EE} \frac{\p }{\p E}\right) g_m. 
\end{equation}
Adding the two equations we get
\begin{equation}
F = D_E~g_m  -  D_{EE} \frac{\p g}{\p E}. 
\label{eq:flux}
\end{equation}

By inserting the values of $D_E$ and $D_{EE}$ into ~(\ref{eq:flux})
we obtain what we will call the fundamental equation for the 
flux
\begin{equation}
\frac{-F} {m \Gamma} = g_m \int_{0}^{E} p g d E +
\left(q \int_E^{\infty} g_m d E + \int_0^E q  g_m d E 
\right)
\frac{\p g}{\p E} 
\label{eq:fun}
\end{equation}
where we have assumed that the zero point of the potential is so chosen so 
that particles have positive energies ($\Gamma = 16 \pi^2 G^2 \ln \Lambda$).

The flux is a radial vector in an energy-parametrised spherical phase space.
In the context of the conventions adopted here, positive flux, pointing outwards, 
corresponds to particle transport towards higher energies.

Another equation that will be useful is the 
flux of background material only. We have  (by using~\ref{eq:flight}) 
\begin{equation}
\frac{-F_\mu} {m \Gamma} =  
\left( q \int_E^{\infty} g_m d E + \int_0^E q  g_m d E 
\right) 
\frac{\p g_\mu}{\p E}, 
\label{eq:flun}
\end{equation}
which does not have a dynamical friction term, and thus invariably 
corresponds to mass outflow into higher energies levels.

\subsection{Linearity of the fundamental equation}
\label{sec:linfun}

Now note that {\em Eq.~(\ref{eq:fun}) is linear in $g_m$}. That is, if a set 
of 
distribution functions $g^l_{m_l} (E)= m_l f^l (E)$ are solutions, any function
\begin{equation}
g_m = \sum_l  A_l g^l_{m_l} (E),
\end{equation}
is also a solution. Therefore one can construct solutions for the various components of 
a multimass system and add them up, assigning arbitrary weights, possibly derived from
assumptions concerning  the effects of mass segregation and stripping 
(cf. Section~\ref{sec:strip}).
Thus our focus on two component systems does not lead to  any fundamental limitation
from the point of view of finding equilibrium solutions.

This remarkable property is a derivative of  the fact that the source term for 
the dynamical friction depends only on the total mass distribution; while the heating,
though originating only from the  massive clumps, affects all particles.
Note, nevertheless, that the evolution of components of  different mass will not be the same,
 since the dynamical 
friction flux for 
{\em each} individual heavy species $i$ would be
\begin{equation}
\frac{F^i_{DF}} {\Gamma} = 
- A_i m_i g_{m_i}  \int_{0}^{E} p g d E, 
\label{eq:fum1}
\end{equation}
while the flux due to the fluctuating force on members of that species is 
\begin{equation}
\frac{F^i_{\rm fluc}} {\Gamma} =
\sum_l A_l  m_l \left( q \int_E^{\infty} g^l_{m_l} d E +  \int_0^E q  g^l_{m_l} d E 
\right)
\frac{\p g_{m_i}}{\p E}. 
\label{eq:fum2}
\end{equation}
Their ratio is $F^i_{DF} /F^i_{\rm fluc}  \sim m_i / \sum_l m_l$.
 Heavier  clumps  will therefore  lose energy to the lighter ones.

From this it follows
that the linearity of the fundamental equation does not imply that 
the evolution of the system can be decomposed into the sum of the evolution of each species 
plus background separately! What it does imply is that, if each distribution corresponding to (e.g) 
distinct mass species happens to be an equilibrium solution for a given total mass distribution, 
the combined system will also be in equilibrium. 
\subsection{Energy change}
\label{sec:energychange}
The amount of 
energy entering the energy surface $E$ per unit time 
is 
\begin{equation}
\frac{d H (< E)}{dt} = E F - \int F d E + E F \frac{\p q}{\p t} - \int_{0}^{E} F \frac{\p q}{\p t} dE.
\label{eq:foken}
\end{equation}
This equation is derived in Appendix~\ref{app:energy} (another derivation  
can  be found in Henon 1961; cf. his Eq~2.27, 2.28 and 4.27)~\footnote{Note that in 
Henon's paper  $S^{'}$
refers to our $F$ and $F$ refers to the distribution function. He 
also uses the term `fundamental equation' to describe the Fokker-Planck 
equation instead of the flux equation, as the term is used here.}.  

The first term of~(\ref{eq:foken}) refers to the energy carried by the mass flux,
while the second represents nonlocal contributions. The last two terms correspond 
to the variation in both these components due to the change 
in phase space volume, which results from change in the potential.
They are analogous to the second term on the right hand side of Eq.~(\ref{eq:fok}).

This form takes into account 
the fact that particles entering into energy surface $E$ from higher energy levels, 
at any given time, do not end up at the same energy $E_f < E$. Neither do they
all originate from the same initial energy level $E_i > E$. A distribution in 
$E_i$ and $E_f$ is always present 
(even if excessively large jumps are unlikely, because of the small deflection 
approximation assumed in deriving the diffusion coefficients). The same goes for 
particles exiting energy surface  $E$ to higher energies. 
Thus, even in the absence of a net mass flux through a 
given surface $E$, energy can still be carried through it. This can lead to 
evolution.

Because of the state of affairs just outlined, 
the evolution driven by the energy flux  will take the form of `heat' transfer 
from  `hotter' regions  of  the system to those that are `colder'. When a temperature gradient 
exists, statistically,  particles crossing $E$ from above will end up at  energies
which, though smaller than $E$, are larger than the compensating particles that crossed 
from below $E$ to exit towards higher energies ---  again, even if the mass flux vanished across
$E$. This is why thermodynamic equilibrium, for a system of solid single mass objects,
can only be isothermal; and all open gravitational systems in virial equilibrium, so composed,
are thermodynamically unstable (e.g., Saslaw, 1985, 2000; Padmanabhan 1990; El-Zant 1998). 
Nevertheless, we will show that in the case of the two component systems considered here, the evolution
timescales  associated with the energy flux can be quite long; and that, under certain
circumstances which may be associated with stripping, even exact equilibria that 
are not isothermal can exist.

\section{Power law solutions}

In this section we consider the case in which both the clump and 
the background distribution functions are power laws in the 
energy. The corresponding physical densities are also 
pure power laws. Perfect power law densities $\rho \sim r^{\gamma}$ 
with $-1.5 \la \gamma \la -1$
are relevant representations in the inner regions
of haloes, where the dynamical friction coupling is most 
efficient. Densities with indices  $-2 \la \gamma \la -1.5$ are useful 
approximate forms at intermediate  
radii, up to the virial radius in haloes of small concentrations
(as discussed in the  penultimate and final  
paragraphs of Section ~\ref{sec:masssol}). We generally have 
in mind a range  $-2 \la \gamma \la 0$, although the case
of $\gamma < -2$, corresponding to the outer regions of haloes 
is briefly considered in Section~\ref{sec:steep}.
Strictly speaking the case of $\gamma = -2$ requires separate treatment, 
because the potential is logarithmic in the radius. Nevertheless,
the flux associated with profiles having $\gamma \rightarrow -2$ continuously tends
to the value at $\gamma = -2$ (namely equilibrium). So will skip such separate
treatment for the sake of brevity.

\label{sec:scalefree}

\subsection{General properties}

\begin{figure}
\psfig{file = 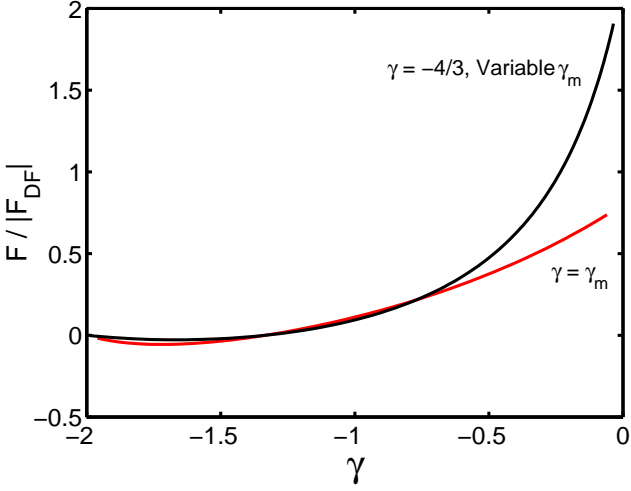,width=92mm}
\caption{Ratio of flux obtained from Eq.~(\ref{eq:fun}), to the 
first term in that equation, describing  the flux of sinking massive 
clumps, for power law distributions  following  the forms given by Eq.~(\ref{eq:g})
and~(\ref{eq:gm}). The corresponding densities are $\rho \sim r^{\gamma}$ for the total
distribution and $\rho_m \sim r^{\gamma_m}$ for the massive clump  distribution.}
\label{fig:scalefree}
\end{figure}

If the physical density can approximated by a power law such that 
\begin{equation}
\rho \sim  r^{\gamma}, 
\end{equation}
the Poisson equation implies a potential is expressed as
\begin{equation}
\phi = \phi_0~r^{- \beta}, 
\end{equation} 
with $\beta =- (\gamma + 2$).
The distribution function that is  
an equilibrium of the unperturbed collisionless system takes the form
(cf., Evans 1994; Evans \& Collett 1997)
\begin{equation}
g = g_0~E^{2/\beta -1/2}.
\label{eq:g}
\end{equation}

Any distribution of a subset of particles 
having a phase space mass density function 
\begin{equation}
g_m = g_{m 0}~E^{2/\beta_m -1/2}
\label{eq:gm}
\end{equation}
has corresponding physical space mass density given by
\begin{equation}
\rho_m =
4 \sqrt{2} \pi \int_0^\infty g_m \sqrt{E - \phi (r)} dE,
\end{equation}
or
\begin{equation}
\rho_m (r) = \frac{ 4 \sqrt{2 \phi_0} g_{m 0} } {2/\beta_m + 1/2}
r^{ - \left(\beta + 2 \frac{\beta}{\beta_m}\right) } \hspace{0.05in} F_1, 
\label{eq:dens}
\end{equation}
where the hypergeometric function
\begin{equation}
F_1 = F_1 (2/\beta_m + 1/2, -1/2; 3/2 + 2/\beta_m; 1) 
\end{equation}
is a constant (that is with third argument equal
to unity)  when both $g$ and $g_m$ are perfect power laws in the energy. The
density of that component is then still a power law, with index
\begin{equation}
\gamma_m = - \left(\beta + 2 \frac{\beta}{\beta_m}\right). 
\end{equation}

Finally,  systems with scale free form for their total mass
distribution function have density of states of the form
\begin{equation}
p = \frac{\p q}{\p E} = p_0~E^{1/2 - 3/\beta}.
\label{eq:p}
\end{equation}

\subsection{Mass flux}  
\label{sec:masssol}

   A steady state implies that all 
time derivatives must vanish. In terms of 
Eq.~(\ref{eq:fok}) this in turn  requires that 
$F \rightarrow 0$.

   Substituting~(\ref{eq:g}) and~(\ref{eq:gm}) into~(\ref{eq:fun}) 
one obtains
\begin{equation}
F = \Gamma m~g_0~g_{m0}~C E^{\frac{3}{\beta_m} - \frac{2}{\beta} + \frac{1}{2}},
\label{eq:sfsol}
\end{equation}
where 
\begin{equation}
C = \frac{- \beta}{\beta -1} + \left(\frac{4 - \beta}{3 \beta - 6}\right) 
\left(\frac{2 \beta_m}{4 + \beta} - \frac{\beta \beta_m}{2 \beta \beta_m - 3 
\beta_m + 2 \beta}\right).
\label{eq:C}
\end{equation}
When $\beta = \beta_m$ Eq.~(\ref{eq:sfsol}) with $F = 0$ reduces to the form 
studied by 
Evans \& Collett (1997), thus their solution with $\beta = -2/3$ and $\gamma = -4/3$ is relevant 
to clump-background systems considered here. 
As noted by these authors, the  only other physical solution, that is a single power law, 
is the singular isothermal sphere ($\gamma \rightarrow -2$).

When the assumption  $\beta =  \beta_m$ is relaxed, there is another 
solutions for each 
value of $\beta$. It however corresponds to  $\beta_m > \beta$.
In the central regions (i.e, where $E,~r \rightarrow 0$), where  
the dynamical friction coupling is strongest, and which must therefore be
the focus of our analysis, the  physical clump
density would be larger than the total; which is unphysical.

 As an example, we plot in  Fig.~\ref{fig:scalefree} the flux obtained from 
Eq.~(\ref{eq:sfsol}), normalised by the
 dynamical friction flux (i.e., by the first term of on the right hand side of that equation
with $C$ substituted in explicitly),
for the   case of $\beta \ne \beta_m$  with $\beta = -2/3$ --- 
an exact solution at  $\beta = \beta_m$.
In addition to this solution, there is another with $\beta_m = -1/2$, 
that is $\gamma_m = -2$.  As $E,~r \rightarrow 0$,  this inevitably implies that
the clump mass density is larger than the total (clump and background combined).
This is clearly impossible (unless a low energy cutoff  is introduced; see below).

 As one moves towards flatter clump density distributions relative to the total,
i.e. $\gamma_m > -4/3$, there is a net positive flux that increases sharply 
beyond $\gamma_m \ga -1$. Because this flux is positive, and increases outwards, there is 
a decrease in phase space density; 
an expansion of the system.  Moreover, this decrease in density is largest as one moves
to smaller energy --- suggesting that initial evolution  proceeds in the direction of  
flatter density distribution. By inserting $g_m = g - g_\mu$,  in Eq.~(\ref{eq:fun}), with $g$ still given by~(\ref{eq:g}) 
and $g_\mu =  g_{\mu 0} E^{2/\beta_\mu -1/2}$, one can arrive  at a 
complementary result concerning the background component. This time, because of the negative sign, one 
concludes that if the background component has a power law distribution that is less steep than 
the total, there is compression of the total  distribution. 
  These deductions are only suggestive; neither rigorous nor comprehensive. They 
are nevertheless supported by the Monte Carlo experiments of  El-Zant, Shlosman \& Hoffman (2001), 
which show that systems with initially homogeneous clump distributions expand.

Predicting the initial direction of evolution in the case when $\gamma = \gamma_m \ne (-4/3, -2)$
is even less straightforward than the case when $\gamma = -4/3$ and $\gamma_m > -1$ discussed above,
 because the flux  diverges at small energies. It would appear that the particle 
distribution is rapidly depleted at these energies, destroying the cusp and forming a core;
while the distribution  steepens outside that (expanding) region (because $\p F / \p E <  0$).
In practice, a small core in the  spatial clump distribution must be always assumed to exist, 
because subhaloes have finite size. It corresponds to a low energy cutoff in the phase space
distribution function (Section~\ref{sec:core}), which would ensure well behaved forms for  the 
flux at $E \rightarrow 0$. Such a small energy cutoff in the clump distribution does not lead to appreciable 
deterioration in the accuracy of the solutions --- as we show in Section~\ref{sec:strip} in connection with the 
issue of stripping.

    An important feature of Fig.~\ref{fig:scalefree} is that {\em although
 for $\beta = \beta_m$ the only exact solution (other than the singular 
isothermal sphere) is that with exponent $\gamma = -4/3$, 
for $\gamma \la -1$ the normalised  flux is quite modest, 
suggesting that these configurations may be long lived.}
In order to make such  statements more precise, 
we consider, in Section~\ref{sec:times},  the question of evolution timescales. They determine
how small the flux needs to be in order for a given approximate solution to be 
considered relevant for the timescales of interest.

Power law densities with  $\gamma \la -3/2$ are steeper than 
inner cusps seen in numerical simulations. They can  nevertheless be retained as 
good represntations to the density profiles of cosmological haloes over a large radial range.
 For example, as can be seen from Fig.~1 of El-Zant \& Shlosman (2002), 
fits to cosmological halo density profiles correspond to softened $1/r^2$ distributions over 
more than three orders of magnitudes in the softening length. These profiles do not incorporate 
faithful reproductions of the phase space structure of the inner cusp, since they lack the 
required temperature inversion in velocity space, but they should represent excellent approximations
at intermediate scales (on which velocity profiles are approximately
isothermal), up to the virial radius for haloes of small concentrations.

Thus, the existence of approximate solutions in the range $-2 \le \gamma  \la  -1$
may  be interpreted to mean  that gravitational systems with density profiles 
similar to those of cosmological haloes are approximately invariant under substructure-induced 
interaction  --- since  dynamical friction coupling at 
$r \gg r_s$ should be negligible, except for the  most massive satellites. 
Nevertheless, as we note below, if $r \gg r_s$, and power laws with $\gamma \ga -2$ no longer  represent reasonable 
approximations to the density distribution, the normalised flux can be large, even 
divergent. This, in principle, may have an effect if dynamical friction coupling on some satellites 
is not entirely negligible, and if a significant fraction of the host halo mass is contained in these regions.

\subsection{Distributions with $\gamma < -2$}
\label{sec:steep}

\begin{figure}
\psfig{file = 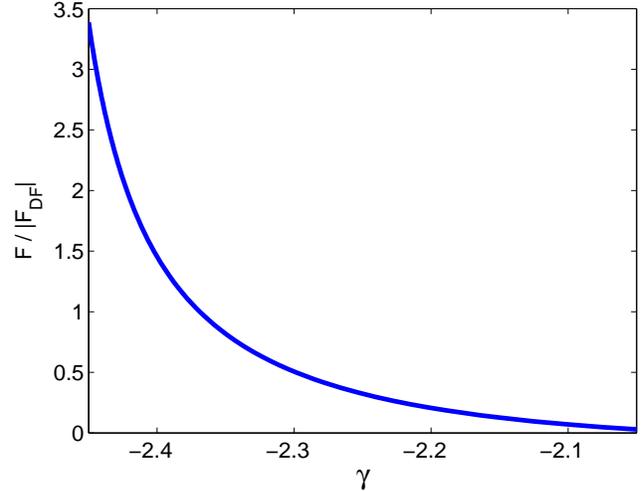,width=94mm}
\caption{Same as in  Fig.~\ref{fig:scalefree}, but for $\gamma < -2$
(Note that Eq.~\ref{eq:fun} has to be slightly modified in this case, 
as explained in text).
The flux diverges for $\gamma \le -2.5$. 
}
\label{fig:steep}
\end{figure}

Although power laws with indices $\gamma \sim -2$ may be useful representations 
of the density  up to the virial radii for cosmological haloes of small concentration,
beyond a few scale length $r_s$, $\gamma \rightarrow -3$.

For $\gamma < -2$, it is appropriate to replace the lower bounds in the first and third integral 
in Eq.~(\ref{eq:fun}) by $- \infty$, and the upper bound in the second integral by $0$. 
Also $E \rightarrow - E$ in equations~(\ref{eq:g}), (\ref{eq:gm}) and  (\ref{eq:p}).

The normalised flux is shown in Fig.~\ref{fig:steep} in the range $-2.5 < \gamma < -2$. 
Only in the immediate neighbourhood of the singular isothermal sphere is the flux 
comparable to the case of $-2 \la \gamma \la -1$. It diverges for $\gamma \le -2.5$.
The source of the blow up is the third integral in~(\ref{eq:fun}). When the total mass
distribution is a power law, so that $q \sim p~E$, it can be seen to correspond (in absolute value) 
to a mass weighed average of the energy of the clump distribution (recall that $M (E)=  p g$).
The excess of low energy clumps causes their average energy to diverge for 
$\gamma \le - 2.5$. The result is a divergent positive flux. For $\gamma \le -3$ the 
first integral also diverges (because the mass does).

In practice, a cutoff is introduced, because $\gamma$ increases as $r \rightarrow  r_s$. 
The severity of the net flux will then depend on the mass inside the region where $\gamma$ falls
significantly below $-2$. Cosmological haloes do not generally contain a large fraction 
of their mass in regions beyond a few scale length $r_s$. The dynamical friction coupling is also exceedingly 
small in these low density regions.  Nevertheless,  the possibility that evolution
on a fraction of a dynamical friction time occurs in the  outer regions of highly concentrated
haloes cannot be completely ruled out within the context of the present analysis; neither can the 
direction  of any prospective evolution be determined.

\subsection{Energy flux}
\label{sec:energysol}
A  thermodynamically stable steady state  requires that 
the time derivatives in Eq.~(\ref{eq:foken}) vanish. 
This in turn implies that $(EF - \int F dE) \rightarrow 0$.
For power law systems  this requirement translates to
\begin{equation}
D = C~\left(1 - \frac{1}{\frac{3}{\beta_m} - \frac{2}{\beta} + \frac{3}{2}}\right)  = 0
\end{equation}
(where $C$ is given by~\ref{eq:C}).
For $\beta = \beta_m$, this is impossible, except for $\beta \rightarrow -2$, corresponding 
to (an unphysical) completely homogeneous system, or $\beta \rightarrow 0$ (the isothermal
sphere). All other solutions are thermodynamically unstable. 
To determine whether these are nevertheless long lived
for the timescales of interest we must derive associated evolution
times. This is done in the next section.

But, for $\beta \ne \beta_m$, 
there are  exact solutions that are 
thermodynamically stable for all values of $\beta$, provided that
\begin{equation}
\beta_m = \frac{6~\beta}{4- \beta}.
\end{equation}
That is,  the massive particle distribution is slightly less steep 
than that of the total. This cannot correspond to any evolution 
with solid clumps --- since, in this case, 
the clumps sink in, replacing the background to dominate the mass distribution
near the centre of the system ($r, E \rightarrow 0$).
It may be relevant however in the presence of 
stripping (Section~\ref{sec:strip}), a process whereby clump material is 
transformed into background. Note that, because these configurations
have $\beta_m \approx \beta$, the corresponding mass flux is also quite
small.

\section{Evolution timescales}

\label{sec:times}

\subsection{Timescales associated with mass flux}

A standard procedure in physics is to derive a characteristic 
timescale by freezing the physical parameters of a 
system at a given moment and estimating the evolution time from there.
In our case, using Eq.~(\ref{eq:fok}), a timescale
for the evolution of the total mass distribution because 
of the existence of a non-vanishing mass flux can then 
be defined as
\begin{equation}
\tau_F  = \frac{M (<E)}{|F (E)|},
\label{eq:tauf}
\end{equation}
which is the time interval the flux at energy $E$ takes
to significantly modify the total mass distribution inside that 
energy surface.

Suppose that, within  that energy surface $E$, the mass distribution is 
initially dominated by the smooth
background component. Then the fluctuations in the potential 
resulting in random heating (described by the coefficient $D_{EE}$) 
have negligible effect on the clumps. They therefore sink in under the action of dynamical 
friction on a timescale similar to that given by~(\ref{eq:DFsim}).
A closely related quantity, the time on which the mass in sinking clumps
significantly modify the original  mass in clumps inside energy 
level $E$ is given by
\begin{equation}
\tau_{F_{DF}} = \frac{M_m (<E)}{|F_{DF}|},
\label{eq:taufdf}
\end{equation}
where $F_{DF}$ as usual corresponds to the first term or the right hand side of 
Eq.~(\ref{eq:fun}).

  The dynamical friction time itself can be arrived at by using Eq.~(\ref{eq:DE}),
again neglecting self interaction terms arising from clump-clump heating due to their mutual 
encounters (i.e., the term involving $\< (\Delta E)^2 \>$). One gets
\begin{equation}
\tau_{DF} = \frac{E}{\< \Delta E \>} = - \frac{p E}{D_E} =  \frac{p E g_m}{|F_{DF}|}.  
\end{equation}
But $M_m (E) = p(E) g_m (E)$; so  we have
\begin{equation}
\tau_{DF} = \frac{E M_m (E)}{|F_{DF}|}.
\label{eq:taudf}
\end{equation}
Since $M (< E)$ is generally smaller than $E M(E)$, this timescale is larger 
than that obtained from~(\ref{eq:taufdf}); it involves the effect of the flux on the local mass 
distribution of clumps, instead of its influence on the  distribution inside the phase
space surface bounded by $E$.

 One can also define an analogous timescale 
for the evolution of the total mass distribution. The ratio of these timescales
would be 
\begin{equation}
R = \tau_{DF} / \tau = \frac{M_m (E)}{M (E)} \frac{|F|}{|F_{DF}|}. 
\label{eq:R}
\end{equation}
It determines how rapidly the total mass evolution takes 
place relative to the dynamical friction timescale; which describes how fast 
the distribution of one of the components, namely the clumps, changes.

Suppose now that the initial system consists of single  power law 
 configurations  
(i.e, in terms of the terminology of Section~\ref{sec:scalefree}, $\gamma_m = \gamma$).
The dynamical friction flux for this initial system then is 
\begin{equation}
F_{DF} = \frac{-m \Gamma}{1-1/\beta}  ~p_0~g_0~g_{m0 }~E^{1/\beta + 1/2},
\end{equation}
(obtained by inserting~\ref{eq:g} and~\ref{eq:gm} into~\ref{eq:fun});
in terms of which one gets an  explicit form of the dynamical friction timescale:
\begin{equation}
\tau_{DF} (t=0)  = \frac{1-1/\beta}{m \Gamma g_0 }~E^{1/2-2/\beta}. 
\label{eq:taudfsf}
\end{equation} 
We omit the explicit reference to the initial time in what follows, even thought this 
is implied.

 For such a system to keep its total phase space mass density
$g$ nearly constant  (Eq.~\ref{eq:g} remaining a good approximation),
for timescales of the order of the initial dynamical friction time, it
is necessary that the absolute value of the dimensionless flux
\begin{equation}
F_e = \frac{\tau_{DF}}{\tau} = \frac{g_m (E)}{g (E)} \frac{F}{F_{DF}} =   \frac{g_{m0} (E)}{g_0 (E)} \frac{F}{|F_{DF}|},
\label{eq:fe}
\end{equation}
be significantly smaller than unity. Note that this same result can be 
inferred from~(\ref{eq:tauf})  and~(\ref{eq:taufdf}). 

Note also that, although this equation assumes a power law form for the {\em total}
distribution, the flux  $F$ can correspond to arbitrary distribution
for any of the {\em individual} (i.e, background and clump) components, 
which are necessarily time dependent.

The quantity $F_e$  {\em defines a relative error}, quantifying
the validity of approximate solutions. More precisely, a solution is a valid 
approximation, that is {\em the system's total phase space 
distribution stays approximately invariant for a timescale $t$ 
if~~$|\< F_e \>_t| \frac{t}{\tau_{DF}}  < 1$~for all relevant energies.}
At any given moment one can  define a dimensionless characteristic 
evolution timescale
\begin{equation}
\tau_e = 1 /  |F_e|;
\label{eq:taue}
\end{equation}
it expresses the number of dynamical friction times the total phase space mass
distribution can remain approximately invariant. A system can be considered unstable 
under the action of the coupling induced by the presence of substructure clumps 
if $\tau_e \approx 1$ --- since, in that case, the total mass distribution changes 
if the configuration is evolved long enough for dynamical friction, the  principal
manifestation of such interaction, to be effective.

As an example, the system studied in Section~\ref{sec:sim} must have 
$\< F_e \> \ll 1$ for energies corresponding to radii that are evolved 
over a dynamical friction time or longer.
 
Fig.~\ref{fig:sim} also suggests that the timescale for each of the components
to be significantly modified is comparable to the dynamical friction time.
This is  consistent with the  results in this section. Note however
 that the replacement of background  by clumps does not proceed linearly in time ---
in which case  it would take a time  $\sim \frac{g_{0}} {g_{m0}}~\tau_{DF} (E)$ for  the clumps to 
dominate  inside some energy surface $E$. 
The  somewhat more detailed analysis of Section~\ref{sec:instab}
 shows that the evolution is actually exponential in $\tau_{DF}$
(at least initially). This explains why the population exchange 
is virtually complete on timescales of the order of a dynamical friction
time, even if the mass fraction in clumps is quite small.

 Finally, note that if the mass fraction in heavy clumps is not negligible,
then the {\em effective} dynamical friction flux decreases by a factor of 
$g_{\mu 0}/g_{0}$, because part of that flux is compensated by 
heating due to clump-clump interaction (c.f., Eq.~\ref{eq:fun}). 
The corresponding dynamical friction time increases by an inverse of this factor.
The  effective evolution flux can then be defined as
\begin{equation}
F_{\rm e, eff} = \frac{g_{m0}}{g_{\mu 0}} \frac{F}{|F_{DF}|}.
\label{eq:effflux}
\end{equation}
Unless otherwise stated, we will have in mind  a system where the background dominates, 
so that $F_{\rm e, eff} \approx F_e$, and will be using  Eq.~(\ref{eq:fe}); since
it corresponds to  a definition of the dynamical friction time that is closer to conventional usage
based on the Chandrasekhar formula, which considers only one massive particle in an infinite
background medium (Eq.~\ref{eq:DFsim} is based on that formulation); obviously, 
conversion between $F_e$ and $F_{\rm e, eff}$ is trivial.

\subsection{Timescale associated with energy flux}

The same attitude adopted in the subsection above  allows one to define 
timescales associated  with the energy flux. For example,
the expression analogous to~(\ref{eq:tauf}) is 
\begin{equation}
\tau_E = \frac{E M (<E)}{|EF - \int F dE|}.
\end{equation}
One can also derive a corresponding `evolution flux'
\begin{equation}
F{eE} =  \frac{g_{m0}}{g_0}\hspace{0.3cm}  \frac{EF - \int F dE}{E |F_{DF}|}.
\label{eq:FeE}
\end{equation}

\section{Evolution timescales of approximate scale free solutions}
\label{sec:timescales}

We now apply the results just derived to power law density distributions with 
$\gamma_m = \gamma$ (considered in Section~\ref{sec:scalefree}).

The values of $F/ |F_{DF}|$ in the case of these scale free solutions 
are shown in Fig.~\ref{fig:scalefree}. For $-2 \le  \gamma \le -1$
the absolute value of  this ratio is $\le 0.111$. For models with single power laws 
$g_{m 0} / g_0 < 1$ determines the mass density of clumps relative to the
total mass density. Eq.~(\ref{eq:fe}) then implies that  evolution 
timescale associated with the mass flux (given by Eq.~\ref{eq:taue})  
is necessarily large compared to the dynamical friction time.

For example, for  the mass ratio of  the simulations
of Sec.~\ref{sec:sim},  $g_{m 0} / g_0 = 0.2$, and so  $F_e \sim 2 \%$ in the 
central region, where the evolution takes place and where $\gamma \sim -1$.
This means that when both mass components are distributed in power law density 
configurations with  $-2 \le  \gamma \le -1$,
the total phase space  mass distribution would evolve over 
$\sim 50$ dynamical friction timescales, very large compared to 
timescales of interest (at $r_s$, this  corresponds to $\approx 15000~\tau_D (r_s)$;
which is far larger than the age of cosmological haloes).

When
$\beta = \beta _m$ the energy evolution flux can be written as
\begin{equation}
F_{eE} = \frac{g_{m0}}{g_0} \frac{\beta^2 (2+ \beta)}{(1-\beta)(1-2 \beta) (\beta+4)}.
\label{eq:thermt_scalefree}
\end{equation}
This reaches a maximum of $0.111 \frac{g_{m0}}{g_0}$ for $\gamma = -1$. It is smaller for other
values of $0 \ge \gamma \ge -2$.  Scale free systems in this range of $\gamma$ are therefore also long lived 
from the energy transfer viewpoint --- that is, the time for energy transfer to affect the total distribution is much longer than the 
dynamical friction time. (note that, for $\gamma = -1$,  $F_e =  F_{eE}$, although there appears to be no reason to suppose 
this  not to be entirely coincidental).

The mass evolution flux is not negligible 
when $\gamma \ga -1$ --- that is, for flat initial 
density distribution --- especially if the mass fraction in clumps is not small.
In  this case one must use Eq.~(\ref{eq:effflux}), with $g_{m 0} \approx g_{\mu 0}$;
which, for $\gamma \ga -1$, implies that $F_{\rm e, eff} \approx 1$.
Therefore, a system which starts with a weak central cusp, and a significant
fraction of its mass in clumps,  is unstable if  the coupling 
due to dynamical friction is non-negligible. The same conclusion  applies to the 
evolution driven by the energy  flux when $\gamma \sim -1$.

\section{Segregation instability}
\label{sec:instab}

Initial scale free solutions 
can only be adopted as  zeroth order approximations.  
As the system evolves, the coupling between the two components 
will cause the light background particles to move out of lower energy levels
and be replaced by the massive clumps. In addition, the latter
can lose mass to the background under the action of stripping.
The goal of the remainder  of this paper 
is to show that the {\em total} phase space mass density can nevertheless 
remain nearly constant, approximately keeping the power law form, even when the individual 
distributions are modified by the aforementioned processes.

Eq.~(\ref{eq:fok}), for the background particles alone, can be written as
\begin{equation}
\frac{\p M_\mu (< E)}{\p t} = - F_\mu (E) + g_\mu \frac{\p q}{\p t},
\label{eq:evol}
\end{equation}
where $F_\mu$ is given by Eq.~(\ref{eq:flun}). Suppose now that the 
total phase space distribution is largely unchanged over timescales 
of interest; that is, $F_e \ll 1$,  and so $g$ continues to be 
 expressed in terms of its initial value form, given by
Eq.~(\ref{eq:g}). The second term on the right hand side 
of~(\ref{eq:evol}) is then also negligible.

We assume a generalised 
power law solution for the  background $g_\mu = g_{\mu 0} (t, E) E^{2/\beta - 1/2}$.
This, in itself, is not an approximation, since the  function 
$g_{\mu 0}$ is arbitrary. However, in order to obtain a solution,
we assume, in addition, that for the purpose of evaluating the integrals in~(\ref{eq:flun}),
$g_m$ keeps its original scale free form, given 
by~(\ref{eq:gm}). For the first  integral in Eq.~(\ref{eq:flun}) this is an excellent 
approximation  at any given time $t$ and energy $E$ such that  $t / \tau_{DF} (E) \la 1$; since, 
in that case, the effect of the dynamical friction coupling is just becoming important at energy $E$, and is 
therefore unimportant at larger energies (recall that $\tau_{DF} \sim E^{1/2-2/\beta}$).

For the second integral, one notes that when 
the {\em total} mass distribution remains in scale free form
$q= p E/ (3/2 - 3/\beta)$. This integral then represents a 
mass weighed average of the energy in  massive clumps
(recall that $M(E) = p g$). It will decrease as the the latter
lose energy because of dynamical friction. Nevertheless, if this
occurs principally at low energies, where $\tau_{DF}$  is smaller,
the change is not drastic. More formally, 
integration by parts transforms it into
\begin{equation}
I / (3/2- 3/\beta) = E \int_0^E  p g_m dE - \int_0^E dE \int_0^E p g_m~dE, 
\end{equation} 
where the scale free  form for (Eq.~\ref{eq:p}) remains relevant, 
in accordance with our assumption that the total distribution is practically 
time independent.  Using Eq.~(\ref{eq:mass}), we get
\begin{equation}
I / (3/2- 3/\beta) =M_m (< E)~E - \int_0^E M_m  (< E)~~dE.
\label{eq:flin}
\end{equation}

Mass conservation implies that the  first term in this expression is constant 
if the effect of dynamical friction is small inside $E$. It is  also
always  larger than the second
\begin{equation}
 M_m (< E)~E~\ge \int_0^E M_m  (< E)~~dE,
\label{eq:Ir}
\end{equation}
where the equality corresponds to the case when all the mass is concentrated 
near  $E = 0$  --- rather unlikely  given the assumption that, at energy $E$, 
the effect of dynamical friction is just being felt at some energy $E \ne 0$, 
(it would imply a sharp discontinuity in the $M_m (E)$ profile).
Moreover, $I$ is  always smaller than the first integral in Eq.~(\ref{eq:flun}), 
which as we have seen is virtually constant. In the initial scale free system
its contribution to the total outward flux is a fraction $(2/\beta+1/2)/(2/\beta + 1/2 -2+1/\beta)$,
which is less than $1/3$ for $\beta \ge -1$ (i.e., for $\gamma \le -1$).
This ratio can only become  smaller on timescales for which the inflow of massive particles across 
$E$ does not significantly change their total mass inside $E$, but where  redistribution 
inside this surface renders their phase space density profile steeper (again this follows 
from  Eq.~\ref{eq:flin}).

The  use of the initial scale free form for $g_m$ in evaluating
the integrals on the right hand side  of~(\ref{eq:flun}) is thus justified
when  $t / \tau_{DF} (E) \la 1$. 
Since the total mass inside energy surface $E$
is still unaffected on that timescale, $M_\mu (< E)$ can also 
be be computed using the 
initial scale free distribution.
In this context, Eq.~(\ref{eq:evol}) transforms into
\begin{equation}
\frac{\p g_{0 \mu}}{\p t} = - A~(g_0  -g_{\mu 0})~g_{\mu 0},
\label{eq:aproel}
\end{equation}
where 
\begin{equation}
A =  m \Gamma~a(\beta)~E^{2/\beta - 1/2},
\label{eq:A}
\end{equation} 
with 
\begin{equation}
a (\beta) = (1-1/\beta) \left(\frac{1}{2/\beta + 1/2} - \frac{1}{2-1/\beta}\right) 
\left(\frac{4-\beta}{3 \beta - 6}\right), 
\end{equation}
a parameter of order unity.
Eq.~(\ref{eq:aproel}) can be integrated to give
\begin{equation}
\frac{g_{\mu 0}}{g_{m 0}} = \chi (E)  e^{- g_0 A t}.
\end{equation}
Here we have assumed, as usual, $g_m = g  - g_\mu$ (and so 
$g_{m 0} = g_0 - g_{\mu 0}$). The function $\chi$ is in principle arbitrary,
but is in fact independent of energy when the initial ratio  of light to 
heavy particles has this characteristic.  And so  we have 
\begin{equation}
\frac{g_{\mu 0}}{g_{m 0}} = \left(\frac{g_{\mu 0}}{g_{m 0}}\right)_{t=0} e^{- g_0 A t}.
\label{eq:lin}
\end{equation}

The timescale
\begin{equation}
\tau_s = \frac{1}{g_0 A}, 
\label{eq:taus}  
\end{equation}
determines the interval on which the exchange of populations, associated with the 
massive clump inflow  accompanied by expulsion of background, takes 
place. This  segregation timescale
is of the order of the dynamical
friction time. More explicitly, using Eq.~(\ref{eq:taudfsf}), 
\begin{equation}
\frac{\tau_{DF}}{\tau_s} = a (\beta) (1- 1/\beta).
\label{eq:DFtos}
\end{equation}

\section{Effect of low energy cutoff on the accuracy of equilibrium solutions}

\label{sec:theinvariance}

\subsection{Behavior at low and high  energies} 
\label{sec:behaviour}

The results of the previous section suggest that 
dynamical friction coupling causes
depletion in the distribution of background particles;
first at the smallest energies, and then at progressively higher
energy.

Suppose  that  there is some cutoff energy $E_c$ below which no background 
particles exist. At energies $E \ll E_c$, and still assuming that 
the clumps are indestructible (no merging or striping), 
there are again  equilibrium solutions which are single power laws 
--- this time for a one component system composed of the clumps,  
since one can now rewrite 
the fundamental equation  only 
in terms of the clumps (by putting $g = g_m$ in Eq.~\ref{eq:fun}).  
The dimensionless evolution fluxes (Eq.~\ref{eq:fe} and~\ref{eq:FeE})
are now significantly larger, since $g_{m 0}/ g_0 \rightarrow 1$; the  
associated timescales
correspondingly smaller.  Nevertheless, 
this state will  still evolve on a timescale that is
$\ga 10~\tau_{DF}$, provided $\gamma \la -1$. Only when $\gamma$
is significantly larger, corresponding to flat density profiles, 
will the evolution timescale be comparable to the dynamical 
friction time.

   It is important to note here that the ultimate effect of dynamical 
friction by a smooth background on solid clumps is to produce a self gravitating 
core made of these clumps, with the total density distribution largely unchanged --- 
{\em and not the sinking of the clumps to the centre} as may be expected had
the background been of infinite mass. No matter how small the fraction of mass 
in clumps initially, the ratio will increase exponentially on a dynamical friction 
timescale (as we have shown in the previous section), with the clumps eventually 
coming to dominate  inside some (time dependent) energy level $E = E_c$.
 This puts an effective end to  effects induced by dynamical friction from the 
background, which has been expelled from that region, and to the sinking of 
clumps to the physical centre of the system.

   For $E \gg E_c$ the system would still retain its original distributions
of clump and background particles, the dynamical friction time being too long
for  there to be any effect. The question therefore is what transpires 
at the boundary; the transition region between the two equilibria.

\subsection{Boundary region: sharp versus gradual cutoff}
\label{sec:inout}

\begin{figure*}
\psfig{file = 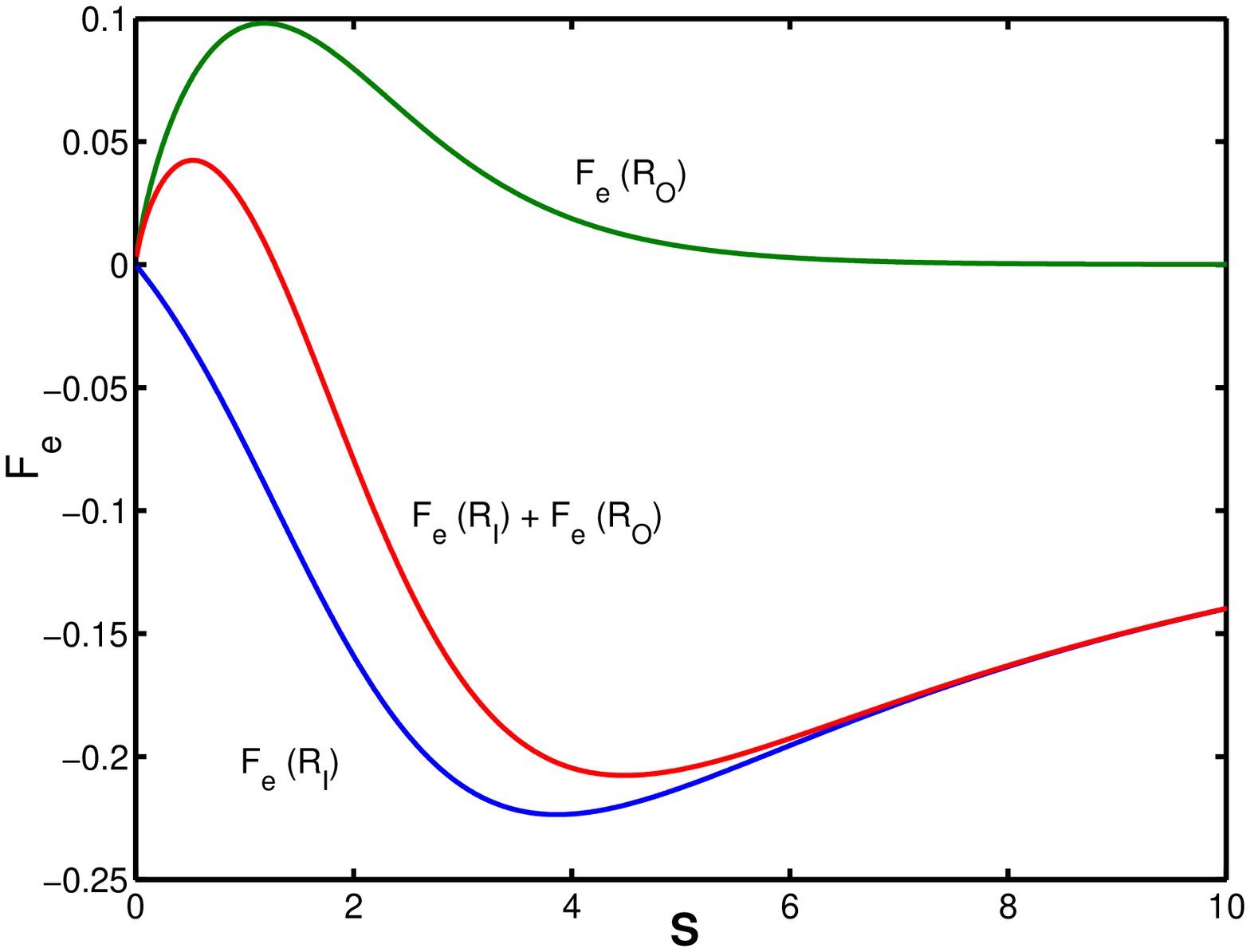,width=87.mm}
\psfig{file = 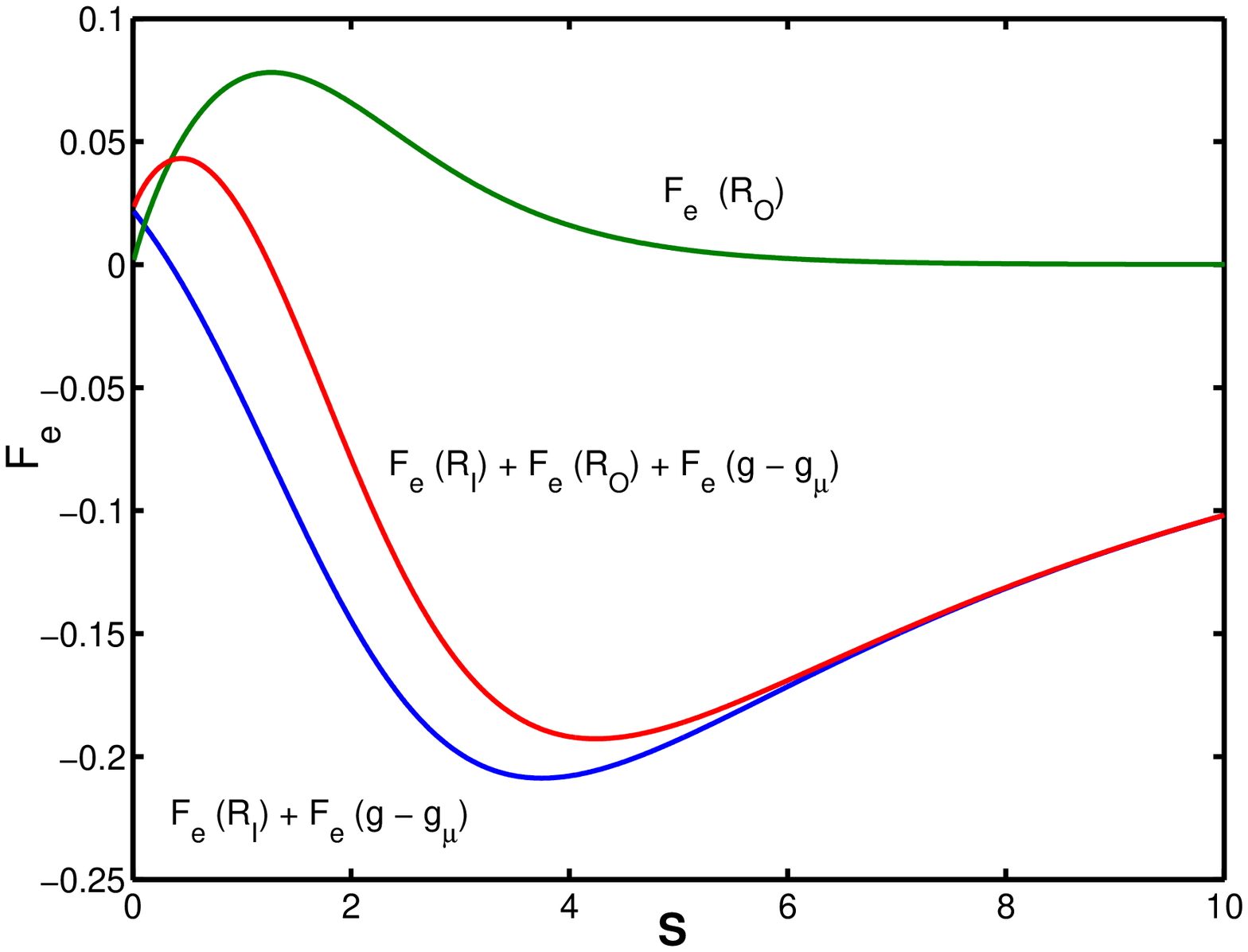,width=87.mm}
\caption{Residual evolution fluxes (calculated using \ref{eq:normresidin} and \ref{eq:normresidout}) 
due to exponential cutoff in the  phase space 
mass distribution function described by Eq~(\ref{eq:lingm0}). Left panel:  $\beta = -2/3$ 
(corresponding to physical density $\rho \sim 1/r^{4/3}$). Right panel:  
$\beta = -1$ ($\rho \sim 1/r$). When $\beta = -1$ we take into account that even systems without the 
low energy cutoff are only in approximate equilibrium (hence the term $F_e (g - g_\mu)$).
The total phase space mass distribution function $g$  is assumed to keep its original 
scale free form (Eq.~\ref{eq:g}).
}
\label{fig:expos}
\end{figure*}

\subsubsection{Mass flux}

  Suppose that  $g_\mu = 0$ for $E < E_c$, but
remains unmodified for larger energies.\footnote{Of course, strictly 
speaking, these conditions cannot be realised simultaneously, since the total mass 
in light particles should be conserved. However the phase space volume 
increases quite steeply with energy ($\sim E^{3/2 - 3/\beta}$), so that 
the migration of low-$E$ particles has little effect on the distribution there
(as in  Fig.~\ref{fig:sim}, where drastic changes in the inner 
distributions of light particles has little effect on the outer regions where they moved 
out). In  what follows we will be  assuming this approximation holds.} 
By writing $g_m = g - g_\mu$ in Eq.~(\ref{eq:fun}) one can separate it into two components:
one involving only functions of $g$ and another involving both $g$ and 
$g_\mu$. If it is  assumed that the total phase space mass 
distribution function $g$ remains a perfect power law, then the 
flux  due to  the  one species term (involving only $g$), 
can be made arbitrarily small ---
by choosing an appropriate value for $\beta$ 
(i.e, $\beta \sim -2/3$ or $\beta \rightarrow 0$). 
The two species component (involving $g$ and $g_\mu$) 
has no such solutions. In fact, for $E < E_c$ there is the residual flux
\begin{equation}
\frac{R_I}{m \Gamma} = \left(q \int_{E_c}^{\infty} g_\mu dE\right) \frac{\p g}{\p E} =
P_0~b(\beta)~E_c^{2/\beta + 1/2} E^{-1/\beta},
\label{eq:rawresidin}
\end{equation}
where $P_0 = p_0~g_0~g_{\mu 0}$ and
\begin{equation}
b (\beta) = - \frac{4 - \beta}{(3 \beta - 6) (2/\beta + 1/2)},
\end{equation}
which is always nonzero. 

Dividing by the dynamical 
friction flux, we obtain the associated evolution flux (cf., Eq.~\ref{eq:fe}) 
\begin{equation}
F_e (R_I)=  \frac{g_{\mu 0}}{g_0} \frac{b (\beta)}{1-1/\beta} \left(\frac{E_c}{E}\right)^{2/\beta+1/2}.
\label{eq:insharp}
\end{equation}
As may be expected from the discussion of the preceding subsection, for $E \ll E_c$
this residual flux becomes  small; the system of clumps, thus represented, is near 
equilibrium. For $E \sim E_c$ 
(and $\beta \sim  -g_{\mu 0}/g \sim -1$) however, it is smaller than 
unity but not negligible.

For $E  > E_c$ there is another non-zero residual term resulting from the 
abrupt cutoff. It is given by
\begin{equation}
\frac{R_O}{m \Gamma} =  -  \left(\int_{0}^{E_c} q g_\mu dE\right) \frac{\p g}{\p E} =
P_0~c (\beta)~E_c^{2 -  1/\beta} E^{2/\beta - 3/2}
\label{eq:rawresidout}
\end{equation} 
with 
\begin{equation}
c (\beta) = -  \frac{4 - \beta}{(3 \beta - 6) (2 - 1/\beta)}.  
\end{equation}
The associated evolution flux this time is 
\begin{equation}
F_e (R_O)=  \frac{g_{\mu 0}}{g_0} \frac{c (\beta)}{1-1/\beta} 
\left(\frac{E_c}{E}\right)^{2 - 1/\beta}.
\end{equation}
It has a sign opposite to that of $F (R_I)$ and  decreases away from $E = E_c$
significantly faster. It is also smaller 
at $E \sim E_c$, but still is not  completely negligible.

Both residual terms can be made far smaller 
if one replaces the abrupt cutoff with a 
gradual one. 
 We do this by introducing  a variable cutoff energy
$E_c = E_c (E)$. This conceptually amounts to dividing the background 
material into a continuum of populations, each with its
own cutoff energy, such that $g_{\mu 0} (E_c)$ changes in
an interval $d E_c$ by an amount $d g_{\mu 0}$. 
We will suppose
that at some energy $E_c = E_c^{\rm min}$, $g_{\mu 0} = 0$; and
at another,  $E_c = E_c^{\rm max}$, it  takes its initial value.
The evolution fluxes can then be written (by summing over all 
the populations)  as 
\begin{equation}
F_e (R_I) = \frac{b (\beta)}{1 - 1/\beta}  
\frac{ \int_{g_{\mu 0}(E)}^{g_{\mu 0} (E_{\rm max})}
E_c^{2/\beta + 1/2} d g_{\mu 0}}
{g_0 E^{2/\beta + 1/2}},
\label{eq:sresidin}
\end{equation}
and
\begin{equation}
F_e (R_O) = \frac{c (\beta)}{1 - 1/\beta}  
\frac{ \int_{0}^{g_{\mu 0} (E)}
E_c^{2 - 1/\beta} d g_{\mu 0}}
{g_0 E^{2 - 1/\beta}}.
\label{eq:sresidout}
\end{equation}

We can also rewrite Eq.~(\ref{eq:sresidin}) and~(\ref{eq:sresidout}) in terms of the dimensionless
variable $X  = E_c / E$. For any phase space point with energy $E$ the `inner flux'
from points with $E_c > E$  is then given by
\begin{equation}
F_e (R_I) = \frac{b (\beta)}{1 - 1/\beta}  
\frac{1}{g_0} \int_{1}^{X_{\rm max}}
X^{2/\beta + 1/2} \left(\frac{\p g_{\mu 0}}{\p X}\right) d X,
\label{eq:normresidin}
\end{equation}
where $X_{\rm max} = E_c^{\rm max}/ E$ and $X_{\rm min} = E_c^{\rm min} / E$.  
There will also be an `outer flux' from points with $E_c < E$. It is
\begin{equation}
F_e (R_O) = \frac{c (\beta)}{1 - 1/\beta}  
\frac{1}{g_0} \int_{X_{\rm min}}^{1}
X^{2 - 1/\beta}   \left(\frac{\p g_{\mu 0}}{\p X}\right)  d X.
\label{eq:normresidout}
\end{equation}
Note that, both these integrals are minimised when  
$g_{\mu 0}$ is flat, that is  
$\frac{1}{g_{\mu 0}}  \left(\frac{\p g_{\mu 0}}{\p X}\right) \ll 1$,
when $g_{\mu 0}$ is largest. By choosing functions with decreasing
logarithmic derivatives, one can make the integrals as small as one 
likes. The fact that the two terms are always nonzero also 
means that they  tend as to cancel each other.
As we will see below, it is easy to find functional forms that 
render the total evolution fluxes quite small.

\subsubsection{Energy flux}
\label{sec:resenflux}

In this case the relevant quantity is $EF - \int F d E$, which
replaces $F$ in calculations analogous to those just described.
It then follows (using~\ref{eq:FeE} and ~\ref{eq:rawresidin})
that 
\begin{equation}
F_{Ee} (R_I) = \left(1- \frac{1}{1-\beta}\right) F_e (R_I)
\end{equation}
and, in an analogous manner (this time using~\ref{eq:rawresidout})
\begin{equation} 
F_{eE} (R_O) = \left(1 - \frac{1}{2/ \beta -1/2}\right) F_e (R_O).
\end{equation}
Thus, for $\beta \sim -1$,
the inner residual energy flux is smaller, and the outer larger, compared to the 
corresponding  mass fluxes  by only a modest factor.  
Since we will be interested in showing that the fluxes are actually smaller by 
an order of magnitude, or more, than those that would lead to significant change
of the total phase space mass distributions, over timescales of interest, 
these factors are of secondary importance. 
In the illustrations that follow we will be focussing on the mass flux.

\section{Accuracy of exponential solutions for the initial evolution}
\begin{figure*}
\psfig{file = 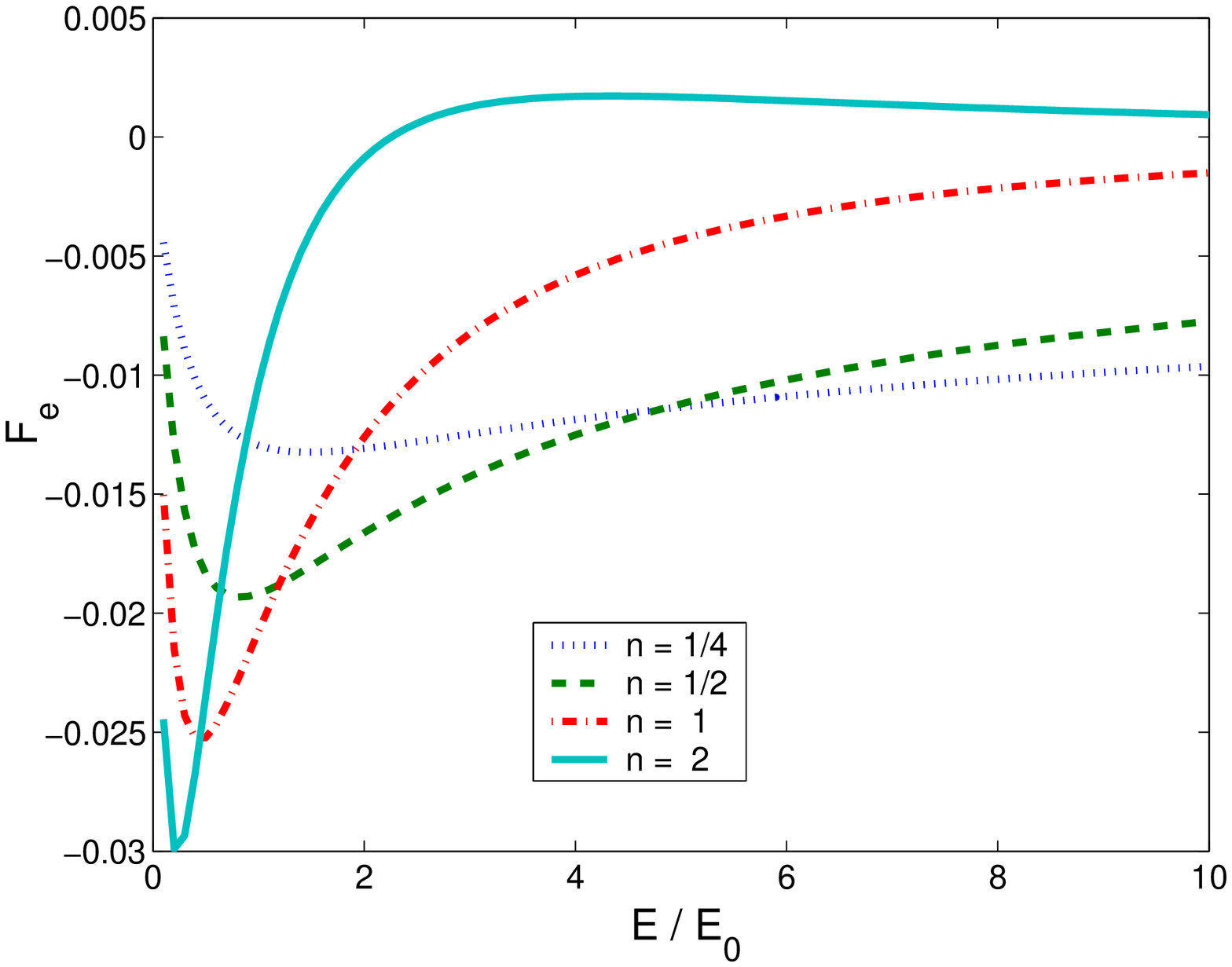,width=84.mm}
\psfig{file = 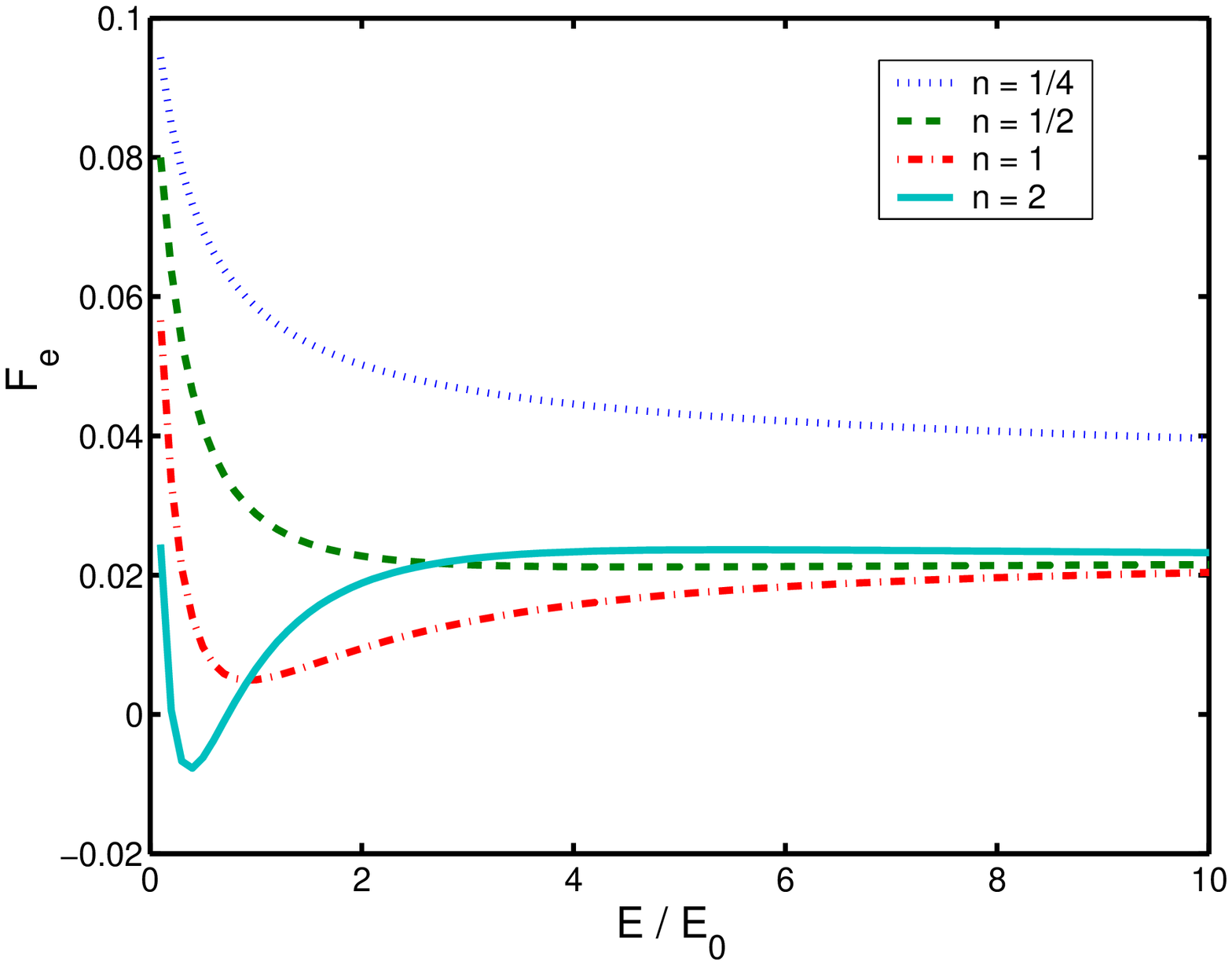,width=91.mm}
\caption{Evolution fluxes (calculated {\em via}  Eq.~\ref{eq:fe}), describing the accuracy of approximate 
solutions of the form (\ref{eq:smooth}), for $\beta = -2/3$ (corresponding to $\rho \sim 1/r^{4/3}$,) shown on the 
left panel, and $\beta = -1$ ($\rho \sim 1/r$) (shown on the right). 
The total phase space mass distribution function $g$ is assumed to keep its original 
scale free form (Eq.~\ref{eq:g}).}
\label{fig:smooth}
\end{figure*}

Equation~(\ref{eq:lin}) was arrived at under the assumption that the total phase space mass 
density remains constant during that evolution; that is $\tau_e \gg \tau_s$.
 We now check this. Self consistency requires that when we insert back the solution
into~(\ref{eq:fun}), 
the evolution flux is indeed small.

 We suppose  that each element of the system 
can be evolved according to Eq.~(\ref{eq:lin}).
According to the discussion  of Section~\ref{sec:instab}, this is 
a good approximation  at points where the effect of dynamical friction
is just becoming significant. Nevertheless, the distribution at smaller energies,
where the dynamical friction time is smaller, need not follow this 
form. This affects the calculation of the third integral 
in~(\ref{eq:fun}) and therefore involves a further approximation, even 
at points where $t \la \tau_{DF}$ (note that it does not affect first integral
which depends only on the total distribution).
But since, as also discussed in the aforementioned section, this integral
(denoted by $I$) is insensitive to the precise distribution inside
energy surfaces $E$ for which $t/\tau_{DF} (E) \la 1$, and in any case
does not dominate the flux, this approximation should be valid when calculating 
that flux for points where the dynamical friction is just becoming important.

 Eq.~(\ref{eq:lin}) can be solved for $g_{m 0}$ at any given energy $E$ and time $t$ to give
\begin{equation}
g_{\mu 0} = \frac{\left(\frac{g_{\mu 0}} {g _{m 0}}\right)_{t=0}}{1 +  \left(\frac{g_{\mu 0}} {g _{m 0}}\right)_{t=0} 
\hspace{0.2 cm}  e^{-  t/\tau_s}}  \hspace{0.4 cm} 
e^{-  t /\tau_s } \hspace{0.3 cm} g_0.
\label{eq:lingm0}
\end{equation}
Substituting this form  for  $g_{\mu 0}$, using Eq.~(\ref{eq:taus}) and~(\ref{eq:A}),
in~(\ref{eq:normresidin}) and ~(\ref{eq:normresidout}) 
and (since this functional form does not admit
sharp cutoffs) letting $X_{\rm min} \rightarrow 0$ and 
$X_{\rm max} \rightarrow \infty$, one obtains the residual fluxes.
These are shown in Fig.~\ref{fig:expos} in terms of 
\begin{equation}
s = \left(\frac{t}{\tau_s}\right)_{X = 1}, 
\end{equation}
and  for $(g_{m0}/g_0)_{t=0} =0.2$.

Since the  exponential evolution was deduced under the assumption 
that material beyond our reference point at  $X = 1$ was largely 
unaffected, they are expected to be highly accurate only for $s \la 2$.
For this is the value of $s$ that corresponds to  one dynamical
friction time for $\beta  \sim -1$ (from Eq.~\ref{eq:DFtos}).

 The results of Fig.~\ref{fig:expos}  do
 indeed show that for $s \la 2$, the average 
value of the evolution timescale (the inverse of the evolution flux shown) 
correspond to $\sim 20~\tau_{DF}$ --- which means  that, up to one dynamical
friction time,  the timescale for the total mass to evolve away from its initial
phase space distribution is twenty times longer. 
The approximation breaks down completely when 
 $t/\tau_{DF}~|\< F_e \>_t| \sim 1/2 s~|\< F_e \>_s| \sim 1$,
where the angular brackets refer to averages of $F_e$ over times smaller than $t$.
From Fig.~~\ref{fig:expos}, it may be estimated that the exponential solutions are relevant
for the whole period of evolution that is plotted. They are a good approximation
($\sim 1/2 s~|\< F_e \>_s| \la 0.2$)
up to  $s \sim 4$, corresponding to about two dynamical friction times.

Note that,
when $s$ is very large,  at $X \sim 1$  the system tends
to the  one-component equilibrium state described in Section~\ref{sec:behaviour}.
Therefore, the accuracy of the exponential solution does not continue to 
deteriorate as $s$ increases. However, because the gradient of $g_{\mu 0}$ is quite large, 
the decay in the inner flux approximately follows the sharp cutoff form given by 
Eq.~(\ref{eq:insharp}), and the tendency towards the one component solution is slow.  
But the form given by~(\ref{eq:lingm0}) has been derived under the assumption 
that $s \sim 1$, and is therefore no longer relevant.
In the following, we present approximate solutions that describe a smooth 
transition between the two equilibria discussed in Section~\ref{sec:behaviour},
with corresponding evolution fluxes that are very small at all points.

\section{Smooth solid clump solutions valid for $\sim 10-100~\tau_{DF}$}

\label{sec:approxal}
\begin{figure*}
\psfig{file = 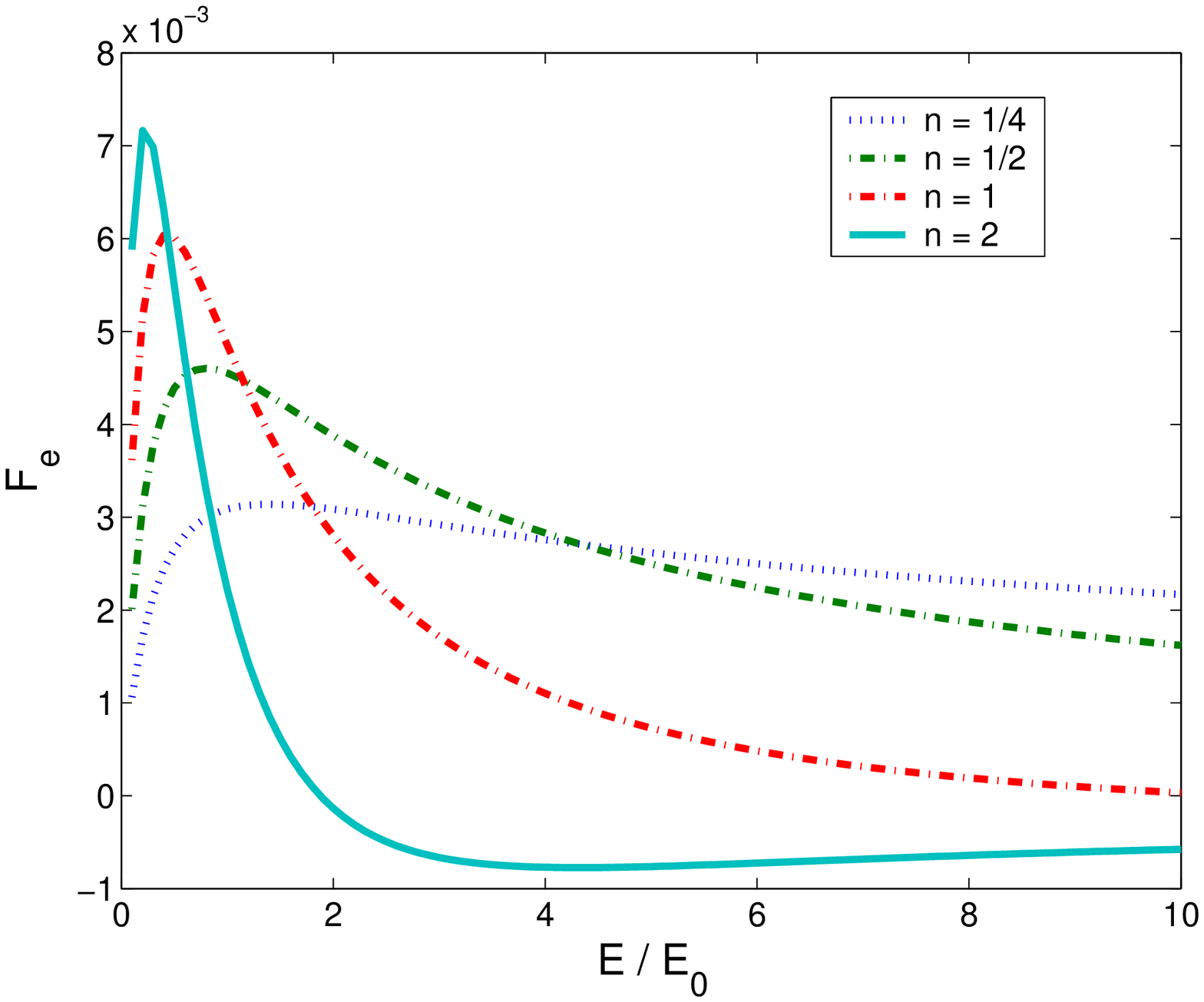,width=87.mm}
\psfig{file = 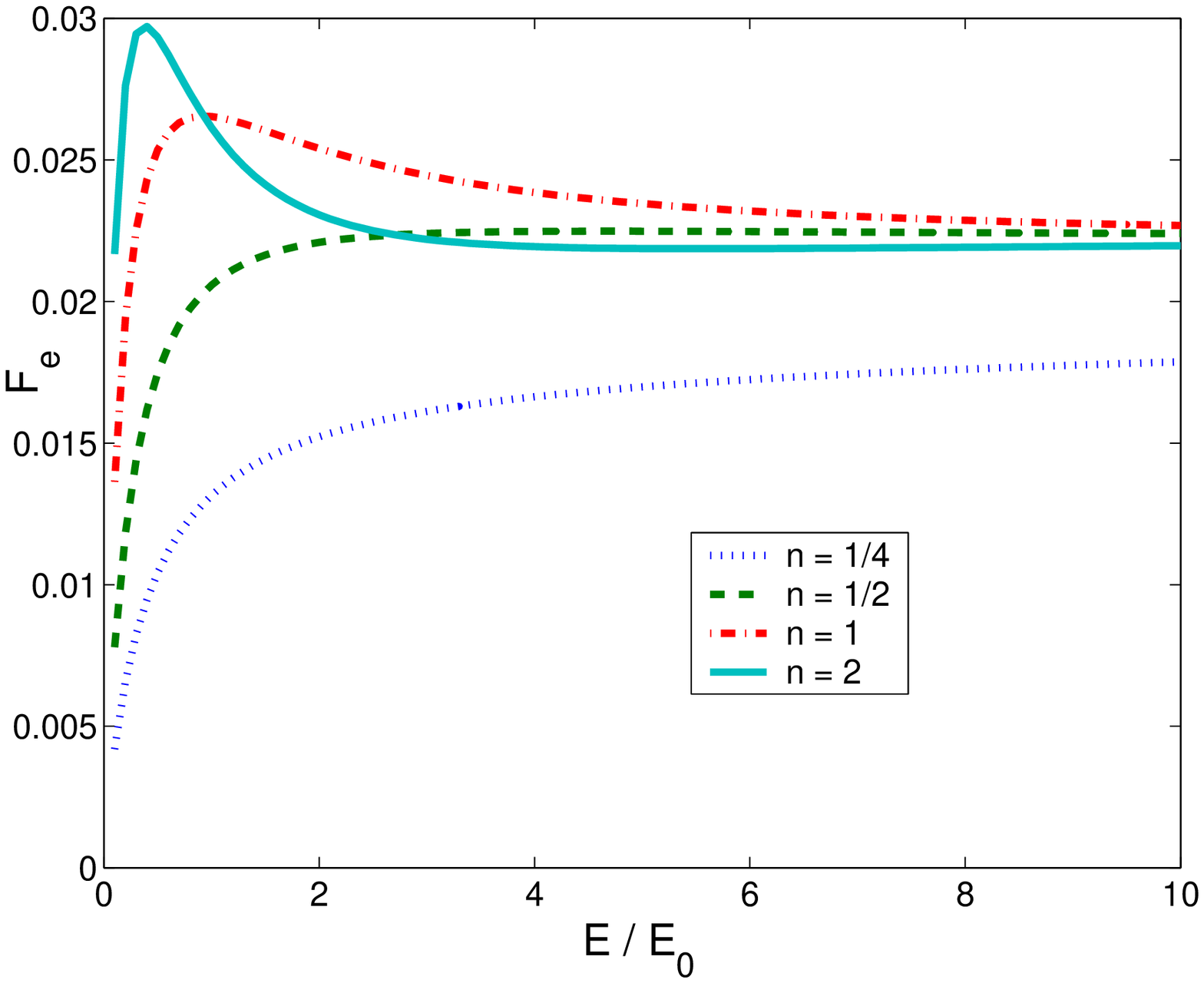,width=87.mm}
\caption{Same as in Fig.~\ref{fig:smooth} but for  approximate 
solutions of the form~(\ref{eq:strip}).}
\label{fig:stripsmooth}
\end{figure*}

Since, for power law forms of the total mass distribution function $g$, 
there only corresponds two solutions for the individual components that 
can be  also expressed in terms of powers of the energy (Section~\ref{sec:scalefree}),
no solutions  can be written as an infinite  power series in the 
energy, and  be exact.  Nevertheless, as we show here, some such solutions can be 
exceptionally  excellent approximations.

For example, we have tried solutions of the form~\footnote{We have also tried other
algebraic (e.g., $\sim (1 + (E / E_0)^n)^{-1}$) and smooth exponential forms 
(e.g., $\sim (1 -  e^{- k E/E_c})$, for $E > E_c$) with similar results.} 
\begin{equation}
g_{\mu 0} (E) = g_{\mu 0} (E_\infty) \left(1 - \frac{1}{(1 + E / E_0)^n}\right), 
\label{eq:smooth}
\end{equation}
where $E_0$ is some scaling parameter and $ g_{\mu 0} (E_\infty)$ 
corresponds to the unperturbed value (i.e. to $g_{\mu 0} (t=0)$).
They represent systems where the background material is depleted
at small energies due to interaction with solid massive clumps.

We insert $g_m = g - g_\mu$
into Eq.~(\ref{eq:fun}), with $g$, the total phase space mass density assumed 
to remain constant.
Consistency requires that the evolution  flux (Eq.~\ref{eq:fe})
is  small.
This substitution breaks  the equation into two additive terms.
In the case when $\beta = -2/3$, the  $g-g$ terms  have exactly vanishing total 
flux. 
For $\beta = -1$ there is an evolution flux of $0.111$. It will add to 
any evolution flux resulting from the $g-g_\mu$ term.

The total  evolution fluxes are shown in Fig.~\ref{fig:smooth} (again for the case
where $(g_{m0}/g_0)_{t=0} = 0.2$).
As may be expected from the discussion in Section~\ref{sec:behaviour},
when $\beta = -2/3$, they  tend  to zero as $E / E_0  \ll  1$
(corresponding to the one component equilibrium state);
and as $E / E_0 \gg 1$ (where  the initial, two component equilibrium, 
still holds). As may also be expected (this time from Section~\ref{sec:theinvariance}),
the absolute values of the fluxes  peak at 
the transition between these two regimes, and are larger when the transition is most rapid.
Note nevertheless that $E_0$ evolves on the local dynamical friction timescale, as the 
instability front moves out to higher energies on the segregation time scale $\tau_s$ of 
Section~\ref{sec:instab}. Therefore  no given region in energy space
would remain, during the whole period of evolution, in the regime where the peaks lie.
Like in the previous section where we discussed the initial exponential evolution, 
it is the average value of the flux that is relevant.

 When $\beta = -1$, and $E / E_0 \ll 1$, the residual flux connected to the $g-g$ (one component) 
term causes the total flux to tend to $0.111$.
The associated evolution timescale is  of the order of 
ten   dynamical friction times.  
The energy flux also becomes important on such timescales. 
Significant evolution in the total mass distribution is therefore expected. 
We have verified by means of simulations (like those of Section~\ref{sec:sim})
that evolution in the total distribution does indeed occur 
 on these  timescales. It is initially
in the direction of expanding 
core (as found by Hayashi et al. 2003). We  have not integrated our systems for 
longer times, but  core collapse, driven by the evolution 
of the (by then completely) self gravitating  system of clumps, probably occurs. 
Both these regimes however are well beyond any situation 
where the representation of substructure in terms  of solid clumps can be expected to be 
of any use. For stripping would have radically modified the distribution and internal 
structure of satellite haloes.

\section{The formation of a constant density core}
\label{sec:core}

The existence of a low energy cutoff in the phase space distribution 
 will entail the removal of some 
particles of the affected species from the central region of the system in physical
space; it does not however imply the removal of all particles. 
Instead, the radial mass density distribution of the 
background will develop a nearly constant density core.
For, as can be seen by replacing the lower integration limit 
by $E_c$ in Eq.~(\ref{eq:dens}), there will always be a radius where 
$E_c \gg \phi (r)$. 
For power law potentials of the form $\phi = \phi_0 r^{- \beta}$,
the transition radius is $r_t = \phi_0 / E_c$. Deep inside this 
region the density tends to the constant value
$\rho_c = 4 \sqrt{2} \pi \frac{g_{m 0} 
E_c^{2/\beta + 1/2}}{2/\beta + 1/2}$.

  This is of course what is found for the background distribution 
in Fig.~\ref{fig:sim} (left panel), where the density of background 
particles does indeed flatten off at small radii. 
In El-Zant, Shlosman \& Hoffman (2001) and El-Zant et al. (2004) we 
had interpreted the clumpy component to be made of baryons; and assumed 
that  stripping was negligible  and that this state of affairs actually 
materialised in practice (the rationale being that baryonic  
material has formed by dissipation and is therefore dense enough 
to survive).

In contrast, if the clumps  are made of dark matter, there will be continuous 
transformation between the two components, as stripping transforms the 
clump material into background. Cosmological simulations actually suggest that 
this process may be so effective that {\em the cutoff in the energy distribution 
in practice probably occurs in the clump distribution}, and not the smooth 
background. For the density distribution of  the clumps consistently
shows evolution towards a flattened core (e.g., Ghigna et al. 2000, Fig.~10).

\section{Stripping}
\label{sec:strip}

As discussed in Section~\ref{sec:striptease},
stripping does not significantly change the phase space mass density 
--- because stripped  particles are likely 
to escape from the clumps with very small kinetic energy and near zero 
binding energy (relative to the clump),  
 the amount of mass in the phase space energy range 
$[E, E + d E]$ remaining nearly constant (the mass in the clumps
in that energy range decreases, but is compensated by an equal increase in 
background).

Therefore, if the interaction due to dynamical friction leaves the 
system in equilibrium, it will remain so  in the presence of tidal stripping.
One nevertheless still needs to show that satellite distributions that are 
affected by stripping also correspond to long lived equilibrium states for 
the total mass distribution.

Suppose now the stripping turns most of the mass in clumps  into background 
materials for energies $E \la E_0$. Then the role of the two components are 
reversed from what was described in Section~\ref{sec:approxal}.
 Now, as appears to be the situation in evolved cosmological systems, 
 it is the satellite distribution which would have a lower energy cutoff and 
develop a core (in physical space), while the background would have a cusp.
In that case, instead of substituting $g_m = g - g_\mu$ into Eq.~(\ref{eq:fun}),
as we have done above, we can directly put
\begin{equation}
g_{m 0} (E) = \frac{g_{m 0} (E_\infty)}{(1 + E / E_0)^n}. 
\label{eq:strip}
\end{equation}
Since this  simply represents a direct exchange of the role of the two 
components compared to what is described by Eq.~(\ref{eq:smooth}), we expect the situation
to be similar  --- except that the evolution 
flux is now expected 
to be even smaller, because the  principal source for evolution, the massive 
clump distribution,
vanishes at small energies, where the coupling is strongest. As can be seen 
from Fig.~\ref{fig:stripsmooth}, this is indeed the case.

Since there is now virtually no 
clump-clump interaction at small energy, there is no longer a relatively large flux 
as $E \rightarrow 0$ when $\beta =-1$. The energy transfer flux is also always small. 
Physically this simply means that, if stripping is efficient over timescales
smaller than the segregation time $\tau_s$ (Eq.~\ref{eq:taus}), the evolved regions tends towards 
a non-evolving collisionless state as $E_0$ increases, instead of the self 
gravitating system of satellites expected to transpire if the stripping time is arbitrarily long
(i.e., when the clumps can be considered solid).

 Of course, in a realistic system, the distribution of clumps would be more complicated. 
There will be a mass spectrum, and each species may have a different 
distribution. Nonetheless, because of the linearity of the fundamental equation (\ref{eq:fun})
with respect to $g_m$ (cf. Section~\ref{sec:linfun}), one can add any number of approximate 
solutions of the form (\ref{eq:strip})
(or, for that matter,~\ref{eq:smooth}) and still end up with a system that is close
equilibrium --- the errors, that is the evolution fluxes, will simply add 
linearly.

\section{Conclusion}
\label{sec:conc}

\subsection{Principal results}

  The principal conclusion of this paper is that the phase space 
mass distribution associated with density distributions akin to the 
 `universal' halo profiles are approximately invariant under the action of the interaction induced 
by the presence of substructure satellites; which is shown to lead to negligible evolution 
over timescales comparable to halo lifetimes. This conclusion applies to the central region ( $\rho \sim 1/r$),
where the dynamical friction coupling is strongest, and up to radii where the profile 
can be approximated as $\rho \sim 1/r^2$.

Since haloes in cosmological simulations build up from smaller components; 
are continually incorporating  infalling material; and
retain significant substructure throughout their evolution, this is a necessary 
condition for the observed universal forms, apparently present at all 
redshifts and mass ranges --- even if the initial collapse naturally leads to 
the `right' distribution, evolution driven by the presence of substructure will modify it, 
unless the total mass distribution is shown to be invariant over relevant timescales.

 The physical mechanism behind our claim is simple. Clumps and background are 
coupled via dynamical friction interaction. The clumps move in, the lighter background 
particles move out, the total remains the same. This equation, which we have shown in 
terms of distribution functions in energy space, naturally leads to invariance in physical space.
It is unmodified by the effect of inter-substructure interaction in the form of weak encounters.

We presented simple simulations which showed that for dozens of dynamical times 
(inside the initial scale length $r_s$), the physical as well as velocity space 
total distributions remain invariant under the dynamical friction mediated 
interaction between a (solid) clumpy component and a background of far lighter 
particles --- despite the fact that, during that same timescale, the distribution
of each component is radically modified (section~\ref{sec:sim}).

 The rest of the paper had the two principal goals: 1) to explain the results of these
simple simulations and 2) To discuss, both from the qualitative and quantitative viewpoints,
the relevance of such idealised models  to the more complex context of cosmological 
simulations.

The conceptual and computational framework employed is based on a Fokker-Planck
formulation, dividing self gravitating structures into two components: a background made of 
light particles and a system of clumps representing substructure satellites. 
The latter are assumed to interact with the background {\em via} dynamical friction,
and among themselves through weak gravitational encounters. Although we 
usually have in mind a system where the background dominates, this is not
an {\em a priori} assumption of the model.

It shows (Section~\ref{sec:scalefree}) that there are exact scale free solutions for cusps 
with both components having scale free density distributions 
$\rho \sim r^\gamma$, and $\gamma = -4/3$. This solution, already found 
for single mass systems by Evans \& Collett (1997), is generalised  for 
the case of  clump-background  systems considered here. 
As long as the massive clumps are much heavier than the background 
particles, any mass distribution of the former is allowed; by virtue of the linearity 
of the fundamental equation for the mass flux (Sec~\ref{sec:linfun}).

Power law  distributions, in the whole range of $-1 \ga  \gamma \ga -2$,  
are in approximate equilibrium, both in terms of the mass and energy 
transfer; the total mass distribution in phase space remaining constant for many dynamical
friction times (Sec.~\ref{sec:timescales}). Solutions with $\gamma \sim  -2$ do not 
correspond to any power law found in the central regions of cosmological 
haloes. But they  are faithful representations of the phase space structure 
at intermediate radii (up to the virial radius for haloes of small concentration;
Section~\ref{sec:scalefree}). Nevertheless, when $r \gg r_s$ and $\gamma \ga -2$
power laws can no longer be taken as faithful represntations of the density distribution,
there can, in principle, be a large mass flux towards higher energies, if dynamical 
friction is not completely negligible at these radii, and if they contain a significant 
fraction of the halo mass (Section~\ref{sec:steep}).

   Even when the total mass distribution function is constant, there is 
continuous evolution in the distribution of each of the components. Two 
cases may be distinguished, depending on how efficient the stripping of 
the massive clumps is.

In the case  when the clumps  are considered as solid objects, we show that the  evolution 
of each of the components  away from the  initial power
law function takes the form of an exponential instability, which we termed the 
`segregation instability', and which takes place on a characteristic timescale comparable
to the local dynamical friction time. It involves the replacement of 
lighter particles, which gain energy and exit lower energy levels, by the massive 
clumps (Section~\ref{sec:instab}).

 This results in a situation whereby  the distribution of light background particles 
develops a low energy cutoff in energy space --- with a corresponding constant density 
core in physical space (Section~\ref{sec:core}), in line with the evolution of the idealised $N$-body
models discussed in Section~\ref{sec:sim}. Long lived solutions for such models are given in 
Section~\ref{sec:approxal}. Situations whereas substructure is formed by dissipationless collapse are unlikely 
to be represented by such models for any significant period of time. Nevertheless, these may be 
relevant if the clumps are made of significantly denser baryonic material; which is more resistant
to dissolution {\em via stripping}, because it dissipated during collapse (e.g. Gao et al. 2004b).   
One then recovers the situation described by El-Zant et al. (2004), where a core developed 
in the light dark matter particles but the total density distribution remained largely 
unchanged. Similar conclusions were reached by Gao et al (2004a) in a much more sophisticated 
cosmological setting.

If the timescale for mass loss from clumps --- stripping --- is significantly smaller
than the segregation time, a cutoff in the {\em satellite} distribution function may result
instead, with an accompanying constant density core in the physical space distribution 
of that component. This appears to be what happens in the context of dissipationless 
cosmological simulations.

 In Section~\ref{sec:striptease} we had argued that  stripping does not modify the total 
mass distribution function. Therefore, if the dynamical friction 
interaction between clumps and background, and weak encounters between 
clumps, do not modify that distribution function, it suffices to show 
that solutions that take into account stripping --- that is ones 
incorporating a low energy cutoff in the clump distribution --- 
can be  long lived. Such solutions, valid up to a thousand dynamical friction times 
(much longer than  the age of cosmological haloes), are given in Section~\ref{sec:strip}.

Finally, it is interesting to note that, whereas in the case of a single mass
system, or a multimass systems without stripping, the only solutions 
with no energy transfer (i.e., thermodynamically stable configuration), 
are isothermal spheres, the case may be different 
when there is a mass spectrum and stripping is 
effective~(Section~\ref{sec:energysol}).

\subsection{Concluding comments}

 Several issues raised in this paper would seem to require 
numerical work, either by means of the Fokker-Planck 
equation or direct simulations, to be settled in a satisfactory manner.
The two approaches are complementary: the full simulations 
including  fewer physical assumptions and the Fokker-Planck models corresponding 
to more controlled calculations that are computationally far less intensive.
The question concerning
statistical effects arising from the finite number of subhaloes, and
the contention that stripping does not modify the total mass distribution 
function (both discussed in Section~\ref{sec:physical}),
need to be empirically verified by means of full simulations, involving
live $N$-body satellites. Other issues are best treated by a combination 
of the two approaches.

The effects of a mass spectrum and 
stripping may  be explicitly included in Fokker-Planck models (Merritt 1983).
It is also possible to incorporate a prescription for the evolution of the halo
mass distribution --- as was done for example by Nusser \& Sheth (1999). These 
authors used a stable clustering formulation, supplemented by an algorithm for
the effect of dynamical friction on substracture and associated back reaction on the main halo.
Because they have assumed that energy lost by sinking satellites is redistributed
homogeneously among parent halo particles, their conclusion was that sinking substructure
causes a steepening of the cusp, invalidating the `universality' assumption. 
Their basic model nevertheless remains relevant and can be coupled to the 
Fokker-Planck formulation, which provides a much better representation of the 
effect of energy deposition by substructure (which in fact is better approximated as a 
local phenomenon).

Of prime interest is  the issue of stability, touched upon  in the introduction --- 
whether the effectively dissipational interaction due to the presence of substructure
may lead to an `attractor' like behaviour, with systems tending to preferred 
configurations similar to those observed in the large simulations. There are several 
related points that are raised by the material presented here.

The fundamental conclusion of this  paper, 
the invariance of the total phase space mass density distribution
under the action of the dynamical friction coupling, strictly speaking
holds only when {\em both components} initially have the same phase 
space distributions. Simple considerations involving the direction of 
phase space fluxes and their gradients (Section~\ref{sec:scalefree}) suggest that
power law systems may expand if the 
the clumps distribution rises less steeply than the background, 
and contract if the situation is reversed. What is available in terms of 
numerical data  tends to confirm this. For example,  
El-Zant, Hoffman \& Shlosman (2001) 
ran Monte Carlo simulations where solid substructure was given  
homogeneous spatial initial conditions inside an NFW halo; the combined 
system was started from virial equilibrium and left to relax towards
detailed dynamical equilibrium; the dynamical friction, modelled on the 
Chandrasekhar formula,  was then turned on. The total density distribution 
decreased. Heuristically, one may explain this from the fact that the binding 
energy available, per unit mass, is larger the more diffuse distribution. 
Nevertheless, this issue is of sufficient importance so as to merit further work, 
involving a series of simulations whereby the spatial distribution of substructure
within the halo, as well as the internal structure of individual clumps is varied.

The material in  Sections~\ref{sec:scalefree} and~\ref{sec:timescales}  suggest that single power law systems with 
flat cusp ($\gamma \ga -1$) are only marginally unstable --- they would evolve on a dynamical 
friction time only when $\gamma \rightarrow 0$ (which may be taken to represent
the nearly homogeneous centre of a system) and when the 
mass fraction in clumps is comparable to that in the background. Whether any evolution actually
occurs will clearly depend on if stripping is effective enough in order to quench the evolution, by modifying 
the mass ratio and cutting the source of the flux before any actual  evolution takes place. 
It is of interest to determine under what (if any) condition evolution does occur, and whether this 
happens in the direction of the cusps characterisitc of cosmological haloes.

Finally, there is the issue of whether the large evolution flux associated with the 
very outer parts of haloes ($\gamma < -2$; see Section~\ref{sec:steep}) has any significance
in determining the structure of the outer profiles. Dealing with these regions
may also require the incorporation of velocity anisotropies.

\section*{ACKNOWLEDGMENTS}
The author is grateful to 
Peter Goldreich for suggestions, inspiration 
and encouragement. It is a pleasure to thank
Adi Nusser and Sergei Shandarin for helpful 
discussions and critique. Comments by 
Julio Navarro were also useful.
This work originated from numerous
discussions with Peter Goldreich, Milos Milosavljevic and 
Isaac Shlosman; it  was supported by NSF 
grant AST 00-98301 at Caltech.

\appendix

\section{Equation for mass change}
\label{app:mass}
In general, a kinetic equation with a collisional term can be written 
as
\begin{equation}
\frac{d g}{d t} = \left(\frac{\p g}{\p t}\right)_{\rm coll}
\label{eq:kinetic}
\end{equation}
where $g$ is mass density distribution function --- the amount of mass 
inside a unit volume element in a six dimensional phase space.
When the collisional term on the right hand side is zero, one recovers 
the collisionless Bolzmann equation:
the dynamics conseerves the phase space mass density along the motion. 

The collisional term can be written as
\begin{equation}
\left(\frac{\p g}{\p t}\right)_{\rm coll} = - \nabla_p F_p,
\label{eq:collp}
\end{equation}
where $F_p$ is the mass flux through phase space, defined as the amount of mass 
crossing unit area per unit time,
and $\nabla_p$ is the phase space gradient. 
This is simply a statment of conservation of mass in phase space --- 
the change in mass density inside a phase space volume due to the effect of 
encounters is the a difference in the amount of mass entering and leaving 
as a result of these encounters. 

For a system where the distribution function depends only on energy, that is 
one whose phase space distribution is spherically symmetric, one can 
rewrite~(\ref{eq:collp}) as
\begin{equation}
\left(\frac{\p g}{\p t}\right)_{\rm coll} = - \frac{1}{p} \frac{\p F}{\p E}.
\label{eq:coll}
\end{equation}
Here $F$ is the mass flux travesing the energy surface $E$. Since this does not
have unit area, one divides by the phase space area of this surface, hence 
the factor $1/p$.  

The term on the left hand side of~(\ref{eq:kinetic}) can be written as
\begin{equation}
\frac{d g}{d t} = \frac{\p g}{\p t} - {\bf V} . \nabla_p g,
\end{equation}
which is  the collisionless Boltzman equation written in terms of the {\em phase space}
velocity ${\bf V}$ of a particle. When the phase space
 is spherical that  velocity is $\dot{E} =  \frac{1}{p} \frac{\p q}{\p t}$; so that
we have
\begin{equation}
\frac{d g}{d t} = \frac{\p g}{\p t} - \frac{1}{p} \frac{\p q}{\p t} \frac{\p g}{\p E}.
\end{equation}
Inserting this into~(\ref{eq:kinetic}), taking into account~(\ref{eq:coll}),
one gets
\begin{equation}
\frac{\p g}{\p t} =   - \frac{1}{p} \frac{\p F}{\p E}   + \frac{1}{p} \frac{\p q}{\p t} \frac{\p g}{\p E}.
\label{eq:FPFC}
\end{equation}
This can be rewritten, in terms of $M (E) = p g$, as
\begin{equation}
\frac{\p M (E)}{\p t} =   - \frac{1}{p} \frac{\p F}{\p E} +  \frac{\p}{\p E} \left(g \frac{\p q}{\p t}\right),
\end{equation}
which, upon integration, gives 
\begin{equation}
\frac{\p M (<E)}{\p t} = - (F (E) - F(0)) + F(t) +  \frac{\p}{\p E} \left(g \frac{\p q}{\p t}\right).
\end{equation}

 For an isolated system in quasi-equilibrium the arbitrary flux $F(t)=0$.
If there are no source or sinks of mass at the centre of the system then also $F(0) = 0$. This is 
the assumption
we adopt here. We nevertheless note that, strictly speaking,
this cannot  not be true of the  approximate scale free solutions presented in Section~\ref{sec:scalefree};
since, for these cases, the flux diverges as $E \rightarrow 0$. One therefore
has to assume a small core that breaks the scale free solutions at the centre. 
In practice this is  naturally realised when the finite size of the clumps is taken into account.
Under the zero central flux assumption  we have
\begin{equation}
\frac{\p M (<E)}{\p t} = - F (E)   +  g \frac{\p q}{\p t}.
\label{eq:dmass}
\end{equation}
  
\section{Equation for energy change}
\label{app:energy}
The rate of energy change of a population of particles in an interval $[E, E+ dE]$ can be related to the 
rate of mass change inside  that interval by
\begin{equation}
\frac{\p H(E)}{\p t} = E~\frac{\p M(E)}{\p t};
\end{equation}
or, in terms of the total mass and energy inside surface $E$, 
\begin{equation}
\frac{\p}{\p E}~\frac{\p H(<E)}{\p t} = E~\frac{\p}{\p E}~\frac{\p M (< E)}{\p t}.
\label{eq:H1}
\end{equation}
Integration (by parts on the right hand side and setting arbitrary time dependent functions and 
central fluxes to zero) 
gives
\begin{equation}
\frac{\p H(<E)}{\p t} = E~\frac{\p M (< E)}{\p t} - \int  \frac{\p M (< E)}{\p t} d E,
\end{equation}
which, when used in conjunction with~(\ref{eq:dmass}) yields the required energy change.

Note that, in  moving from the local energy $H (E)$ to the integrated one $H (<E)$,
we have neglected the interaction potential of the particles, that is we have simply
added their energy. This is consistent with the context of the Fokker-Planck formulation,
since the assumption of locality in relation to the encounters producing the energy
changes (c.f. Section~\ref{sec:FP}) already incorporates the idea that the 
evolution is cuased 
by encounters which proceed independenly of the form of the mean field, which is  
responsible for this interaction. This is no longer true if singificant energy changes are caused 
by an evolving mean field. However, for our purposes, this point is not of central importance; 
the rationale being that we assume that the system is in quasiquilibrium and evolves only due
to the encounters. The source of any evolution is 
then the Fokker-Planck flux. 

\section{Diffusion coefficients}
\label{app:diff}
In general, diffusion coefficients represent the averge rates that a certain 
quantity changes over time. In case of the isotropic orbit averaged 
Fokker-Planck equation they represent the average rate of change of powers of the 
specific energy $E$, averaged all orbits with energies in the interval
$[E, E + dE]$. (Note that, as always with this equation, this involves the assumption
that the actual rate is such that the change during an orbital time is small, thus permiting the 
averaging of quantities over orbits).

In the weak encounter (Fokker-Planck) approximation, two diffusion coeficients are 
relavant: $D_1 = \<\Delta E \>$ and $D_2 = \< (\Delta E)^2 \>$. For a test particle of 
mass $m_t$ affected by encounters with field particles of mass $m_f$ and distribution 
function $f_f$, they are given by (Spitzer 1987; Theuns 1996)
\begin{equation}
D_1 = \Gamma m_f^2~\left(\int_E^{\infty} f_f~dE - \frac{m_t}{p m_f} \int_0^E q~f_f~dE\right)
\label{eq:D1}
\end{equation}
and 
\begin{equation}
D_2 = 2~\Gamma m_f^2~\left( \frac{1}{p} \int_0^E q~f_f~dE + \frac{q}{p} \int_E^\infty f_f~dE\right)
\label{eq:D2}
\end{equation}
where $\Gamma = 16 \pi^2 G^2 \ln \Lambda$, $\Lambda$ being the Coulomb logarithm; $q$ is given by~(\ref{eq:q})
 and $p=\frac{\p q}{d E}$ (note that in this paper we are assuming that the zero of the potential 
is taken so that energies are always positive).

 In a system consisting of several mass species, field particles will include all other particles of the 
same species, plus those of the other species present. The form of the diffusion coeficients is additive in the 
distribution of these different species, it is thus easy to generalise relations obtained for a two 
component system. We therefore consider such a system; with  constituents having  masses $m$ and $\mu$.
The phase space {\em mass} distribution functions are $g_m  = m f_m$ and $g_\mu = \mu f_\mu$.
We will assume that the phase space mass densities of the two components are comparable; that is 
$g_m$ and $g_\mu$ are roughly of the same order, while at the same time $m \gg \mu$.     
For the average energy change of the light particles one therefore has
\begin{equation}
\< \Delta E_\mu \> = - \Gamma m \int_E^\infty g_m~dE,
\label{eq:deltEmu}
\end{equation} 
while for the massive species it is 
\begin{equation}
\< \Delta E_m \> = \Gamma m \left(\int_E^\infty g_m~dE
- \frac{m}{p}  \int_0^E p g~dE\right),
\label{eq:deltE}
\end{equation}
where $g = g_m + g_\mu$.

The average square changes in specific energy 
 are 
\begin{equation}
\< (\Delta E)^2 \> = 2~\Gamma m \left(\frac{1}{p} \int_0^E q g_m~dE + \frac{q}{p} \int_E^\infty g_m~dE\right).
\label{eq:deltEE}
\end{equation} 
This relation holds  {\em for both} the light and massive  species.

\end{document}